**Improved proteomic analysis of nuclear proteins, as exemplified by the comparison of two myeloïd cell lines nuclear proteomes**

**Cécile Lelong[(1)*], Mireille Chevallet[(2)], Hélène Diemer[(3)], Sylvie Luche[(1)], Alain Van Dorsselaer[(3)] and Thierry Rabilloud[(4)]**

[(1)] Laboratoire de Chimie et Biologie des Métaux, Université Joseph Fourier-Grenoble, iRTSV/LCBM, CEA Grenoble, 17 avenue des Martyrs, 38054 Grenoble cedex 9, France

[(2)] Laboratoire de Chimie et Biologie des Métaux, iRTSV/LCBM, CEA Grenoble, 17 avenue des Martyrs, 38054 Grenoble cedex 9, France

[(3)] Laboratoire de Spectrométrie de Masse Bio-Organique, Département des Sciences Analytiques, Institut Pluridisciplinaire Hubert Curien, UMR 7178 (CNRS-UdS), ECPM, 25 rue Becquerel, 67087 Strasbourg Cedex 2, France

[(4)] Laboratoire de Chimie et Biologie des Métaux, UMR CNRS 5249, iRTSV/LCBM, CEA Grenoble, 17 avenue des Martyrs, 38054 Grenoble cedex 9, France

* To whom correspondence should be addressed: cecile.lelong@ujf-grenoble.fr
Tel (++33) (0) 438 783 212
Fax (++33) (0) 438 789 803

Keywords: 2D-gel electrophoresis, dendritic cell, macrophage, nuclear protein extraction

## Abstract

One of the challenges of the proteomic analysis by 2D-gel is to visualize the low abundance proteins, particularly those localized in organelles. An additional problem with nuclear proteins lies in their strong interaction with nuclear acids. Several experimental procedures have been tested to increase, in the nuclear extract, the ratio of nuclear proteins compared to contaminant proteins, and also to obtain reproducible conditions compatible with 2D-gel electrophoresis. The NaCl procedure has been chosen. To test the interest of this procedure, the nuclear protein expression profiles of macrophages and dendritic cells have been compared with a proteomic approach by 2D-gel electrophoresis. Delta 2D software and mass spectrometry analyses have allowed pointing out some proteins of interest. We have chosen some of them, involved in transcriptional regulation and/or chromatin structure for further validations. The immunoblotting experiments have shown that most of observed changes are due to post-translational modifications, thereby a exemplifying the interest of the 2D gel approach. Finally, this approach allowed us to reach not only high abundance nuclear proteins but also lower abundance proteins, such as the HP1 proteins and reinforces the interest of using 2DE-gel in proteomics because of its ability to visualize intact proteins with their modifications.

## Introduction

2D-gel technology is a powerful proteomic tool to visualize the expressed proteins of a cell at a defined time and in a defined condition. One way of study is the comparison of proteins differentially expressed in two extracts resulting of the modification of one parameter, e.g. stress condition compared to physiological condition of growth or two stages of differentiation of a cell lineage. Such comparisons may lead to some bias in the proteins highlighted. In point of fact, one of the crucial points in 2D-gel based proteomics is the solubilization of proteins, which has to be adapted to the 2D-gel constraints on one hand, and yet allow to observe a maximum of proteins in an experiment on the other hand. The combination of both conditions leads to the "loss" of certain categories of proteins (e.g. poorly soluble proteins) and the highlighting of some other ones (e.g. most abundant proteins and "Déjà vu" metabolism). In eukaryotic cells, the most abundant and soluble proteins are particularly predominant because of the high number of proteins potentially expressed and above all the compartmentalization of the cell, which prevents the solubilization of certain proteins. This compartmentalization may also be an advantage because it may be a good way to visualize the low abundant proteins presenting a high dynamic range of concentration, or the proteins specific of a metabolism localized in an organelle. Several experimental procedures are now available to specifically separate each type of organelle from the rest of the cell using subcellular fractionation. But it is also well known that these procedures are not perfect and that proteins from other compartments always contaminate an organelle preparation. We are particularly interested to the nuclear proteome to better understand and identify transcription factors and protein regulators that control eukaryotic gene expression involved in cell differentiation. The nuclear proteomes of several eukaryotic organisms, mammary epithelial cells [1], yeast [2] or amniotic epithelial cells [3] have already been studied by 2D-gel. In most cases, an important part of identified proteins (from 30 to 60%) following a differential analysis are not bona fide nuclear proteins [4], even though some papers have reported a better proportion of nuclear proteins [5]. However, many of the nuclear proteins reported using classical extractions are involved in RNA metabolism, and the fraction of the nuclear proteins that interacts with nucleic acids is still under-represented, unless specific enrichment procedures such as DNA affinity chromatography is carried out [5]. To visualize and analyze such proteins, the solubilization procedure has to be adapted. We have tested different ways to solubilize and separate them from DNA, to increase the ratio of nuclear proteins compared to contaminant proteins like cytosolic proteins or proteins from other organelles. As the ultimate goal is to perform comparisons between different biological conditions, the chosen procedure must be able to extract chromatin proteins and still allow obtaining reproducible conditions with several cell types and/or culture growth conditions and be compatible with 2D-gel electrophoresis. The NaCl procedure is the procedure that fits best to all these constraints. We have compared the proteome of the NaCl extract with the proteome of total nucleus extract, and identified a largest proportion of nucleus proteins in the spots specifically highlighted in the NaCl extract (not present in the total nuclear extract).

Then, we have tested this experimental procedure to compare the nuclear protein expression profiles of two cell types, macrophage and dendritic cells by 2D-gel electrophoresis. We compared J774 and XS52 cell lines, which are representative of macrophages and dendritic cells respectively, and are known to present different phenotypes. Using the Delta 2D software, we have selected and identified by mass spectrometry 193 proteins showing significant differences between the two groups of them essentially involved in transcriptional regulation and/or chromatin structure using immunoblotting approaches (in one and two dimensional gel): HMGB1,2 and HP1$\alpha$,$\gamma$ proteins.

## Materials & Methods

### Cells lines

J774 cells were obtained from ATCC, XS52 were obtained from Dr Bernd Kleuser (Marburg, Germany). J774 cells (mouse macrophages) and XS52 cells (mouse dendritic cells Langerhans subtype) were grown in tissue culture flasks (BD Falcon™) in High glucose DMEM (Sigma) supplemented with 10% fetal calf serum (FCS), and 10mM Hepes. All mouse cell cultures were supplemented with ciprofloxacin (30µM final) and maintained in a 37°C incubator in 5% $CO_2$, up to a density of 1 million cells/ml.

**Protein preparations**.

Total protein extract: The cells were rinsed with phosphate-buffer saline and then swollen in one volume of TES buffer (10mM Tris, 1mM EDTA, 0.2M Sucrose). Four volumes of lysis buffer (8.75M urea, 2.5M Thiourea, 5% CHAPS, 12mM Tris carboxyethylphosphine, 25mM Spermine) were added and the proteins extracted for 30 min at room temperature. The lysate was ultracentrifuged at 320,000xg for 30 min at 20°C. The supernatant was then collected and frozen at -20°C.

Nuclei preparation: Nuclei were isolated by modification of a published method [6]. The cells ($10^9$) were rinsed with phosphate-buffer saline and lysed at 0°C in 10 volumes of buffer A (10mM Hepes, pH 7.5, 1mM DTT, 1mM Spermidine, 0.25mM Spermine, 0.5mM EDTA, 10mM KCl and 0.05% Triton X100) for 20 min. 0.2M Sucrose was added to the suspension before centrifugation at 1000xg for 5 min. The pellet, containing the nuclei, was washed in 10 volumes of buffer B (10mM Hepes, pH 7.5, 1mM DTT, 2mM $MgCl_2$, 0.2M Sucrose) and then centrifuged at 1000xg for 5 min at 4°C. The pellet was resuspended in storage buffer (10mM Hepes, pH 7.5, 25% Glycerol, 5mM $MgCl_2$, 0.1mM EDTA, 5mM DTT) and frozen at -20°C.

Total nuclear protein extract: The nuclei pellet was resuspended in extraction buffer (7M urea, 2M thiourea, 4% CHAPS, 10mM Tris carboxyethylphosphine and 20mM spermine base) and then centrifuged at 320,000xg for 30 min at room temperature. The nuclei preparation was quantified using Bradford assay before adding ampholyte (0.4% w/v final).

Benzonase nuclear protein extract (adapted from [7]): 1000u of benzonase and 0.2mM of dicholoroisocoumarin (serine protease inhibitor) were added to 30µl of nuclei suspension. The volume was adjusted to 50µl with $H_2O$. The sample was incubated at 37°C during 30 min. The sample was then diluted in 10mM EDTA, 50mM DTT and 2% SDS in 100µl final volume, boiled for 3-5 min at 100°C and then cooled in a cold water bath. One volume of 5% cresol in water-saturated phenol was added. The sample was vortexed at least five times and centrifuged at 10000xg for 10 min. The phenol phase was collected and dialysed overnight in 0,1% SDS, 5mM Tris-base pH7.5, 0.25M sucrose, 1mM EDTA and 5mM DTT. The proteins were finally concentrated by the TCA/sarkosyl precipitation protocol [8].

DNase nuclear protein extract: The nuclei preparation was diluted in 10 volumes of 0.25M sucrose, 5mM EDTA pH 6.4 and 0.2mM dichloroisocoumarin, incubated 15 min at 4°C. 100u of DNase II was added. The mixed sample was incubated 1h at 37°C and then centrifuged at 10,000xg for 5min. The supernatant was collected and concentrated by the TCA/sarkosyl precipitation method.

Urea-salt nuclear protein extract: One volume of nuclei preparation was diluted in four volumes of 6M urea, 1M NaCl and 20mM spermidine, incubated at room temperature for 1h and centrifuged at 320,000xg for 30min at 20°C. The supernatant was collected and concentrated by the TCA/sarkosyl precipitation method.

NaCl nuclear protein extract: The nuclei preparation was diluted in 10 to 20 volumes of 10mM Tris-Base pH 7.5 and 0.35M NaCl, incubated for 30 min at 4°C and then centrifuged at 320,000g for 30 min at 4°C. The supernatant was collected, diluted three times with cold water and concentrated by the TCA/sarkosyl precipitation method.

NaCl/SB3-12 nuclear protein extract [9]: The nuclei preparation was diluted in 10 to 20 volumes of 10mM Tris-Base pH 7.5, 0.35M NaCl and 1% SB3-12 incubated for 30 min at 4°C and then centrifuged at 320,000g for 30 min at 4°C. The supernatant was collected, diluted three times with cold water and concentrated by the TCA/sarkosyl precipitation method.

Lecithin nuclear protein extract [10]: One volume of nuclei preparation was diluted in four volumes of 0.5% lecithin, 7M urea, 2M thiourea, 0.4% ampholyte and 50mM phosphoric acid-HCl pH 2.8, incubated 1h at 25°C, and centrifuged at 320,000xg for 30min at 20°C. The pellet was resuspended in rehydratation solution without ampholytes (7M urea, 2M thiourea, 4% CHAPS).

Protein quantitation: Total protein extracts, total nuclear protein extracts, DNase nuclear protein extracts and lecithin nuclear protein extracts were quantified using Bradford assay. NaCl nuclear protein extracts were quantified after separation and staining of SDS-PAGE: known quantities (10, 20 and 30 $\mu$g) of total protein extracts and nuclear protein extracts, and two dilutions of the NaCl nuclear protein extracts were separated on the same gel. Proteins were stained with colloidal coomassie blue [11]. The NaCl protein extract concentration was then estimated with the ImageJ freeware.

**2D-gel electrophoresis**

IEF: Home made 160mm long 4-8 or 3-10.5 linear pH gradient [12] gels were cast according to published procedures [13]. Four mm-wide strips were cut, and rehydrated overnight with the sample, diluted in a final volume of 0.6 ml of rehydratation solution (7M urea, 2M thiourea, 4% CHAPS, 04% carrier ampholytes (Pharmalytes 3-10) and 100mM dithiodiethanol [14, 15].

The strips were then placed in a Multiphor plate, and IEF was carried out with the following electrical parameters: 100V for 1 hour, then 300V for 3 hours, then 1000V for 1 hour, then 3400V up to 60-70 kVh. After IEF, the gels were equilibrated for 20 minutes in Tris 125mM, HCl 100mM, SDS 2.5%, glycerol 30% and urea 6M. They were then transferred on top of the SDS gels and sealed with 1% agarose dissolved in Tris 125mM, HCl 100mM, SDS 0.4% and 0.005% (w/v) bromophenol blue.

SDS electrophoresis and protein detection: 10%T gels (160x200x1.5 mm) were used for protein separation. The Tris taurine buffer system was used at a ionic strength of 0.1 and a pH of 7.9 [16]. The final gel composition is thus Tris 180mM, HCl 100mM, acrylamide 10% (w/v), and bisacrylamide 0.27%. The upper electrode buffer is Tris 50mM, Taurine 200mM, SDS 0.1%. The lower electrode buffer is Tris 50mM, glycine 200mM, SDS 0.1%. The gels were run at 25V for 1hour, then 12.5W per gel until the dye front has reached the bottom of the gel. Detection was carried out by fast silver staining [17].

Image analysis: Image analysis was performed with the Delta2D software (v.3.6) (Decodon, Germany). Briefly, 3 gel images arising from 3 different cultures and nuclear preparations were warped for each group onto a master image, one for the J774 cell line and one for the XS52 cell line. The XS52 master gel image was then warped onto the J774 master gel image and a union fusion image of all the gel images was then made. The detection was carried out on this fusion image, and the detection results were then propagated to each individual image.

The resulting quantification table was then analyzed using the Student t-test function of the software, and the spots having both a p-value lesser than 0.05 and an induction/repression ratio of 2 or greater were selected for further analysis by mass spectrometry after all being manually verified. For the global analysis of the power and reproducibility of the experiments, the Storey and Tibshirani approach was used [18] as described by Karp and Lilley [19].

The spots of interest were excised from a silver staining gel by a scalpel blade and transferred to a 96 well microtitration plate. Destaining of the spots was carried out by the ferricyanide-thiosulfate method [20] on the same day as silver staining to improve sequence coverage in the mass spectrometry analysis [21]. In some cases, to maximize sequence coverage and avoid the artefacts associated with silver staining, the ultrafast carbocyanine fluorescent stain was used [22].

**Mass spectrometry analysis**

In gel digestion: In gel digestion was performed with an automated protein digestion system, MassPrep Station (Waters, Milford, USA). The gel plugs were washed twice with 50 µL of 25 mM ammonium hydrogen carbonate ($NH_4HCO_3$) and 50 µL of acetonitrile. The cysteine residues were reduced by 50 µL of 10 mM dithiothreitol at 57°C and alkylated by 50 µL of 55 mM iodoacetamide. After dehydration with acetonitrile, the proteins were cleaved in gel with 10 µL of 12.5 ng/µL of modified porcine trypsin (Promega, Madison, WI, USA) in 25 mM $NH_4HCO_3$. The digestion was performed overnight at room temperature.

MALDI-MS analysis and protein identification: Mass measurements were carried out on an UltraflexTM MALDI-TOF/TOF mass spectrometer (Bruker Daltonics GmbH, Bremen, Germany) under control of Flexcontrol 2.0 software (Bruker Daltonics GmbH, Bremen, Germany). This instrument was used at a maximum accelerating potential of 25 kV in positive mode and was operated in mode reflector at 26 kV. The delay extraction was fixed at 110 ns and the frequency of the laser (nitrogen 337 nm) was set at 20 Hz.
Sample preparation was performed with the dried droplet method using a mixture of 0.5 µl of sample with 0.5 µl of matrix solution dry at room temperature. The matrix solution was prepared from a saturated solution of alpha-cyano-4-hydroxycinnamic acid in water/acetonitrile 50/50 diluted three times in water/acetonitrile 50/50.
The acquisition mass range was set to 400-4000 m/z with a matrix suppression deflection (cut off) set to 500 m/z. The equipment was first externally calibrated with a standard peptide calibration mixture that contained 7 peptides (Bruker Peptide Calibration Standard #206196, Bruker Daltonics GmbH, Bremen, Germany) covering the 1000-3200 m/z range and thereafter every spectrum was internally calibrated using selected signals arising from trypsin autoproteolysis (842.510 m/z, 1045.564 m/z and 2211.105 m/z). Each raw spectrum was opened with flexAnalysis 2.4 (Bruker Daltonics GmbH, Bremen, Germany) software and processed using the following parameters: signal-to-noise threshold of 1, Savitzky-Golay algorithm for smoothing, median algorithm for baseline substraction, and SNAP algorithm for monoisotopic peak detection.
The proteins were identified by peptide mass fingerprinting using a local Mascot server with MASCOT 2.2.0 algorithm (Matrix Science, London, UK) against UniProtKB SwissProt and TrEMBL databases (version 20080905, 6462751 entries). The research was carried out in all species. Spectra were searched with a mass tolerance 50 ppm, allowing a maximum of one trypsin missed cleavage. Carbamidomethylation of cysteine residues and oxidation of methionine residues were specified as variable modifications. Proteins are validated when the ratio of the number of matched peaks on the total number of peaks is higher than 60%, and if i) the position of the spot in the pI dimension was within the theoretical pI ± 1pH unit, and if ii) the position of the spot in the Mw dimension corresponded to at least 90% of the theoretical Mw. This strategy was implemented to remove proteolytic fragments from our protein identifications.

NanoLC-MS/MS analysis and protein identification: NanoLC-MS/MS analysis was performed using an Agilent 1100 series nanoLC-Chip system (Agilent Technologies, Palo Alto, USA) coupled to an HCTplus ion trap (Bruker Daltonics GmbH, Bremen, Germany). The system was fully controlled by ChemStation B.01.03 (Agilent Technologies, Palo Alto, USA) and EsquireControl 5.3 (Bruker Daltonics, Bremen, Germany). The chip was composed of a Zorbax 300SB-C18 (43 mm × 75 μm, with a 5μm particle size) analytical column and a Zorbax 300SB-C18 (40 nL, 5 μm) enrichment column. The solvent system consisted of 2% acetonitrile, 0.1% formic acid in water (solvent A) and 2% water, 0.1% formic acid in acetonitrile (solvent B). The sample was loaded into the enrichment column at a flow rate set to 3.75μL/min with solvent A. Elution of the peptides was performed at a flow rate of 300 nL/min with a 8-40% linear gradient of solvent B in 7 minutes. For tandem MS experiments, the system was operated in Data-Dependent-Acquisition (DDA) mode with automatic switching between MS and MS/MS. The voltage applied to the capillary cap was optimized to -1800V. The MS scanning was performed in the standard/enhanced resolution mode at a scan rate of 8100 m/z per second. The mass range was 250-2500 m/z. The Ion Charge Control was 100000 and the maximum accumulation time was 200 ms. A total of 4 scans was averaged to obtain a MS spectrum and the rolling average was 2. The three most abundant precursor ions with an isolation width of 4 m/z were selected on each MS spectrum for further isolation and fragmentation. The MS/MS scanning was performed in the ultrascan mode at a scan rate of 26000 m/z per second. The mass range was 50-2800 m/z. The Ion Charge Control was 300000. A total of six scans was averaged to obtain an MS/MS spectrum.

Mass data collected during analysis were processed and converted into .mgf files using DataAnalysis 3.3 (Bruker Daltonics GmbH, Bremen, Germany). A maximum number of 250 compounds was detected with an intensity threshold of 60000. MS spectra were smoothed by Savitzky Golay algorithm with a smoothing width of 0.2 m/z in one cycle. A charge deconvolution was applied on the MS full scan and the MS/MS spectra with an abundance cutoff of 5% and 2% respectively and with a maximum charge state of 3 and 2 respectively.

For protein identification, the MS/MS data were interpreted using a local Mascot server with MASCOT 2.2.0 algorithm (Matrix Science, London, UK) against UniProtKB SwissProt and TrEMBL databases (version 20080905, 6462751 entries). The research was carried out in all species. Spectra were searched with a mass tolerance of 0.2 Da in MS and MS/MS modes, allowing a maximum of one trypsin missed cleavage. Carbamidomethylation of cysteine residues and oxidation of methionine residues were specified as variable modifications. Protein identifications were validated when the Mascot protein score was above 60, and if i) the position of the spot in the pI dimension was within the theoretical pI ± 1pH unit, and if ii) the position of the spot in the Mw dimension corresponded to at least 90% of the theoretical Mw. This strategy was implemented to remove proteolytic fragments from our protein identifications.

**Immunoblotting analysis**

Protein concentration was measured using the Bradford assay. 20 μg of each crude extracts were separated on a 10% SDS-PAGE gel and then electrotransferred (Bio-Rad system) onto nitro-cellulose membranes (Bio-Rad). Membranes were blocked with PVP40 1% in PBS-0.1% Tween overnight [23]. They were then probed with appropriate dilution of primary antibodies raised against HP1α (ab64916), or HP1γ (ab10480), or HMGB1 (ab18256), or HMGB2 (ab61169). This was followed by incubation with appropriate horse-radish peroxidase-conjugated secondary antibodies (Abcam or Santa Cruz Biotechnology). The blots were developed with the ECL kit (Amersham Biosciences). To normalize the total protein quantities really transferred, the membranes were stained with 0.5% india ink/ PBS-0.1% Tween overnight [24], and then rinsed twice in PBS-0.1% Tween. The intensity of each immunoblot band and of the total protein transferred was quantified with the ImageJ software. The results are presented as follow: (intensity of the ECL band in XS52 extract/ intensity of the ink total XS52 protein extract) / (intensity of the ECL band revealed in J774 extract/ intensity of the ink J774 total protein extract). Figures 5 and 6 were the results of at least three measurements from three independent cultures of each cell line.

## Results and Discussion

Nuclear protein extractions

To analyse nuclear protein with a 2-DE approach, we have tested different methods to extract proteins from the nuclei. The protein extraction methods must be compatible with 2-DE conditions and efficient to give sufficient quantities of proteins. The first step consists to prepare total nuclear protein extract from a nuclei preparation. From this total nuclear extract, we have tested different ways to enriched the extract in nuclear proteins and separate them from DNA using different agents as urea, DNAse, benzonase, non-ionic detergent (SB-12) or lecithin (Fig. 1). The DNAse and urea extractions led to the poorest yield and IEF (Fig. 1b,c). The benzonase extraction also led to insufficient IEF (Fig. 1d). The NaCl combined to SB3-12 and the lecithin extractions led to a similar protein profile to the NaCl extraction but resulted in an insufficient yield (Fig. 1a,f). These three different extraction conditions showing similar profiles, we have ruled out general effect of TCA on protein proteolysis (which is not used in to prepare the lecithin nuclear protein extract). Moreover, most of the protein identifications (Table 1) were within a MW/pI window corresponding to theoretical data. The NaCl extraction method gave the best results: spots were well focused with a good yield (Fig. 1e). Moreover, this method being one of the simplest with few steps, has constantly given the most reproducible protein profiles. To obtain consistent gels showing around two thousand spots, we have separated 150μg of proteins which correspond to around $2x10^6$ and $0.25x10^9$ cells for total nuclear protein extract and the NaCl protein extract, respectively. The NaCl extract method require at least a hundred fold more cells than the total nuclear protein extract method to obtain gels with an equivalent quality. To test the relevance of such extraction, we must be sure that two main problems were resolved. First, we have to check the quality of the nuclear fractionation knowing that proteins coming from other compartments always pollute subcellular fractionations. Second, we have to analyse the NaCl pattern to be sure that proteins interacting with DNA as chromatin proteins or transcriptional regulator are effectively enriched compared to "pollutant" proteins. To ensure this both goals, we have compared the protein patterns from three extractions methods: the total protein extract (Fig. 2a), the total nuclear protein extract (Fig. 2b) and the NaCl nuclear protein extract (Fig. 2c). On the total protein extract, we have identified, by mass spectrometry, proteins systematically present compared to the nuclear (total or NaCl) extracts. These spots being very abundant because of the very different patterns, we have randomly chosen spots covering most of the area of the gel (Table 1). The main part of identified spots are localized in the cytoplasm (64% corresponding to cytoplasm 17% + cytoplasm and nucleus 47%) (Fig. 2d). 31% of the identified spots are secreted or localized in organelles or in the cell membrane. 5% of the identified spots are exclusively nuclear. In the two nuclear extracts, apart from a few abundant housekeeping proteins, the nuclear protein patterns are very different from the total protein extract, so that a direct comparison is not very meaningful (Fig. 2a compared to 2b and 2c). So, we have identified spots systematically enriched in a nuclear pattern compared to the other one (total nuclear extract against NaCl nuclear extract). The comparison of the two types of nuclear extracts shows that the NaCl extraction promotes the identification of nuclear proteins and moreover, of proteins interacting directly with DNA like the Purβ transcriptional regulator or the hnRP complex. Most of proteins enriched in NaCl nuclear protein extract are localized in the nucleus (52.5 % corresponding to 9.5% in the nucleus and 43% in the cytoplasm and nucleus, Fig. 2f). And the part of protein exclusively localized in the cytoplasm is reduced to 33% compared to 42% in the total nuclear extract (Fig. 2e). Reciprocally, the part of proteins localized both in the cytoplasm and the nucleus, is of 43% in the NaCl nuclear extract against 26% in the total nuclear extract. No protein localized exclusively in the nucleus has been highlighted in the total nuclear extract (Fig. 2e). Thus, despite its lower yield of protein extraction compared to the total nuclear extract, we have decided to use the NaCl nuclear extraction method, with which we have enriched the nuclear part of the extract (52.5% in the NaCl nuclear extract compared to 26% in the total nuclear extract), to compare the nuclear proteomes of different cell lineages. To test this nuclear protein extraction methodology, we have compared the nuclear proteomes of two cell types which have been differentiated from a common myeloid progenitor, the circulating blood monocytes: a dendritic cell line, XS52 and a macrophage cell line, J774. These two cell lines represent two different myeloid cell types (macrophage for J774 and Langerhans cells for XS52), so that we could expect to detect differences in their nuclear proteomes.

Comparison of NaCl protein extracts from macrophage and dendritic cell lines

Three independent samples of NaCl protein extracts from a murine macrophage cell line (J774) and three from a murine dendritic cell line (XS52), i.e. made from three independent cultures, were compared by 2-DE followed by silver staining. Around two thousand spots have been detected on each gel.

Delta 2D software analyses have highlighted 101 spots on 4-8 homemade strips (Fig. 4a) and 92 spots on 3.7-10.5 homemade strips (Fig. 4b), which were differentially expressed by a factor equal or greater than two and a p-value lower than 0.05 in a two-tailed t-test. The t-test distribution analysis showed that we could expect more than 50% true positives when detecting differentially-expressed proteins, while the null experiment (J774 against itself, using independent gels and cultures) showed a much weaker proportion (Fig. 3). All the differentially expressed spots between J774 and XS52 have been identified by mass spectrometry (Table 2). 62.6% of identified spots are nuclear proteins (35% exclusively nuclear and 27.6% with at least one subcellular localization described to be the nucleus) in an equivalent way in the two cell lines: they are involved in transcriptional regulation (COMMD1, COMMD3, Dr1 or Purβ), splicing (hnRPs or ISY1 homolog), (hetero)chromatin modelling (nucleoplasmin, HP1α, HP1γ, HMGB1 or HMGB2) or replication (Rfc2, Rfc4 MCM4 or MCM5). The other identified proteins are described being localized in the cytoplasm (19.5% exclusively in the cytoplasm), in different organelles (e.g. 4% in the mitochondria, 0.8% in the golgi) or in the membrane (10.6% for at least one localization). It is interesting to note that the part of identified proteins localized exclusively in the nucleus and moreover described to interact with DNA represents a significant part of the identified proteins (35%). We have decided to focus on four proteins which are, directly or not, involved in gene expression regulation and for which antibodies were available: HMGB1, HMGB2, HP1α and HP1γ.

HMGB1 and HMGB2

HMG (High Mobility Group) proteins are the second most abundant chromosomal proteins, after histone proteins, and are thought to play important roles in modelling the assembly of chromatin and in regulating gene transcription in higher eukaryotic cells. They play essential roles in a variety of cellular processes such as cancer development, DNA repair and infectious/inflammatory disorders. All HMG proteins are subjected to a number of post-translational modifications, which modulate their interaction with DNA and other proteins. Three distinct families of HMG proteins have been defined and named based on the structure of their DNA-binding domains and their substrate binding specificity: HMGA (HMG-AT hook), HMGB (HMG-box) and HMGN (HMG-nucleosome binding). HMGB family includes HMGB1, HMGB2 and HMGB3. They exhibit different gene expression patterns: HMGB1 is ubiquitous, whereas HMGB2 is primarily expressed in the thymus and testes and HMGB3 expression is localized to the bone marrow [25] [26]. HMGB1 and -2 enhance the binding of various transcription factors like p53 [27] or Rel family proteins [28] [29] [30] [31]. Five spots corresponding to HMGB1 and -2 have been identified on 2-DE analysis: two for HMGB2 (spots 58 and 59, Fig. 4a) and three for HMGB1 (spots 61, 62 and 63, Fig. 4a). This multispot patttern is consistent with what has been already described for HMG1 [32] and HMG2 [33]. This, together with the absence of artifactual modifications of major spots such as actin, suggests that our procedure does not induce artifactual modifications of the proteins.

All of the HMGB spots have an increased intensity value in the NaCl nuclear extract from XS52 cells compared to the NaCl nuclear extract from J774 cells (Table 2). We have performed immnunoblot analysis in one and two dimensions gels to evaluate the relative quantity of each HMGB protein with another approach (Fig. 5). The immunoblot quantitation of HMGB2 in a total protein extract in one or two dimensions has shown that HMGB2 expression level is equivalent in the both cell lines (Fig. 5a,b). The 2-DE immunoblot analysis of HMGB2 spots (Fig. 5b) has revealed that the main part of HMGB2 spots is basic (pI between 7 and 9) and that the pattern and the total intensity are constant. The two spots first identified in 2-DE analysis belong to the acidic part of the pattern. The different modified forms have probably different expression levels relatively to their function but are not representative of the total quantity. By contrast, the immunoblot analysis of the HMGB1 protein has confirmed the observation of the 2-DE analysis: HMGB1 is more expressed in XS52 cells than in J774 cells (ratio of 2) (Fig. 5a). The immunoblot analysis of total protein extract by 2-DE have confirmed this result (Fig. 5b): the patterns were identical whatever the cell lines but the expression level was higher in XS52 cell line in comparison to the pattern of the J774 cell line. Systematic analysis by mass spectrometry of spots of the same MW in the area where the first HMGBs spots have been identified, have shown that at least seven (pI between 6 to 7) and nine spots (pI between 7 to 9) correspond to HMGB1 and -2, respectively (data not shown). In the case of HMGB1 proteins, all spots display the same intensity in a cell line relatively to the other. The PTMs of HMGB1 seem not to be relevant to explain phenotype differences between the two cell lines. Only the relative concentration of HMGB1 proteins differs from a cell line to the other. In contrast, two spots of HMGB2 seem to have a differential pattern of expression in the two cell lines while the total quantity of HMGB2 is constant. These results show that although belonging to the same family of proteins, HMGB1 and HMGB2 are regulated by different mechanisms. When the comparison between XS52 and J774 is made HMGB1 is regulated by a translational mechanism (all protein species are regulated in the same extent) while HMGB2 is regulated by a posttranslational mechanism (only the most acidic i.e. most modified forms are increased).

HP1α and HP1γ

Heterochromatin protein I (HP1), first discovered in Drosophila, is a protein family that is evolutionary conserved, from fungi, to plants and animals. There are multiple members within the same species. HPI proteins are composed of two domains: the amino-terminal chromodomain binds methylated lysine 9 of histone H3, causing transcriptional repression, and the highly conserved carboxy-terminal chromo-shadow domain enables dimerization and also serves as a docking site for protein involved in a wide variety of nuclear functions, from transcription, regulation of euchromatin genes to nuclear architecture [34] [35] for reviews. HP1 proteins are amenable to posttranslational modifications that probably regulate these distinct functions [36] [37]. Takanashi and collaborators [38] have shown that HP1γ decreases during adipocyte differentiation, whereas HP1α and HP1β are constitutively expressed. Three spots corresponding to HP1α and HP1γ were detected on 2-DE: spot 1 (= HP1γ) (Fig.4a) and spot 127 (= HP1γ) and 137 (= HP1α) (Fig.4b). They were all localised in the same small area (Fig. 6b). The intensity ratios (XS52/J774) estimated with Delta 2D software were similar for the both spots identified as HP1γ (XS52/J774 = 0.48 and 0.5) and opposite for HP1α (XS52/J774 = 2) (Table 2). In contrast, in a total protein extract and for the HP1α and HP1γ, the immunoblot quantifications have shown that the intensity levels were identical in the both cell lines (XS52 and J774) (Fig. 6a). This result was confirmed with immunoblot analysis of total protein extract separated by 2-DE (Fig. 6b). These experiments detected several spots for the HP1 proteins: at least three for HP1α and nine for HP1γ (Fig. 6b). Only two of them for HP1γ and one for HP1α have been detected with a modified expression level between J774 and XS52. The different modified forms have probably different expression levels relatively to their function but the total quantity of HP1(γ or α) protein is constant in differentiated cell lines. HP1 is known to be highly posttranslationally modified [36] [37]. Similar expression profiles have already been described during adipocyte differentiation [38]. These results show that the regulation made on HP1 proteins between J774 and XS52 is made essentially at the post translational stages suggesting that modifications play an important role in the modulation of the functions of HP1 proteins.

**Concluding remarks**

One of the major problems of proteomics is undersampling. In 2D gel-based proteomics, this undersampling results in the visualization of a limited number of proteins, so that many studies end up with the same types of proteins [39], that belong to the core stress response of animal cells [40]. In order to reach lower abundance proteins, it is necessary to focus the proteomic analysis to a subcellular subset. A good example is represented by secreted proteins, for which a sensitivity down to 1ng/ml can be reached [8]. When this type of sensitivity is reached, the classical differential proteomic analysis is able to go deeper and reach less common proteins [17].

In this frame, nuclear proteins represent a good way to investigate by proteomics the mechanisms underlying the changes made in gene expression during a biological process. However, nuclear proteins are rather difficult to extract under conditions compatible with proteomics. First of all, nuclei are very rich in DNA, and this DNA must be eliminated. Second, DNA-bound proteins are of great interest when dealing with processes involved in changes in gene expression. However, these proteins are often not extracted by low ionic strength solutions, even in the presence of urea [41]. Conversely, they are easily extracted by salt [42]. A good example is represented by the HMG proteins, which are very abundant chromatin proteins. They are not present when the nuclei are extracted with urea, but are easily detected as soon as salt is used for extraction ([5], this work). However, high concentrations of salts are incompatible with many types of proteomics, including 2D gel-based proteomics. We therefore coupled salt extraction, used to effectively extract DNA-bound proteins in proteomics setups [1] [5] with the TCA-sarkosyl precipitation process, which has proven to be of high yield and devoid of efficiency thresholds [8]. However, we had to dilute the salt-containing sample before precipitation, otherwise the residual salt concentration in the final sample remained too high for 2D gel electrophoresis. Overall, our data show that the NaCl extraction methodology allowed observing nuclear proteins that interact with nucleic acid such as protein Dr1 (proteins associated with transcriptional regulator) or histone proteins (i.e. HMGB proteins). Moreover, the similarity of the patterns observed with two cell lineages that arise from a common myeloid progenitor, show the robustness of this extraction approach. In spite of this similarity, more than a hundred spots have a significantly different expression in one lineage compared to the other one. 35% of this differentially expressed spots are exclusively nuclear proteins.

In fact, this technique offers an interesting extraction of nuclear proteins without the added complexity of the DNA chromatography used in [5]. Moreover, it retains bona fide chromatin proteins that are lost during the DNA chromatography, such as the chromobox proteins.

This approach allowed us to reach not only high abundance nuclear proteins, such as HMGs, but also lower abundance proteins, such as the HP1 proteins, or transcriptional factors, i.e. Purβ. We could easily study proteins that are implicated at the level of chromatin structure, such as HMG and HP1 proteins. Interestingly, the immunoblotting experiments carried out to confirm the results of 2DE-gel unravelled the fact that many detected changes do not correspond to an overall increase in the amount of proteins, but to a change in the PTM pattern of the proteins.

This is another reminder that observing an increased on a 2D gel does not necessarily means that the amount of the total gene product has changed, and control experiments such as 1D or 2D blots are required to demonstrate a real increase in the amount of the protein identified, and not just an increase in a specific form of this protein. This also underlines the possible importance of these PTM in the functions of the proteins and thus in the general control of gene expression. Further studies of these PTM will be needed to understand in more detail the link between these PTM and the modulation of the function of the proteins. This observation reinforces the interest of using 2DE-gel in proteomics because it allows seeing intact proteins. Further identifications of PTM associated with the different phenotypes of myeloïd cells will provide new insight in mechanisms controlling gene expression. This study is the first step of a PTM analysis of nuclear proteins involved in genetic regulations.

Table 1

| Protein Identification | | | | | Prot. Local. (4) | Mass spectrometry analysis (nLC-MS/MS or MALDI/TOF-PMF) | | |
|---|---|---|---|---|---|---|---|---|
| **Spot nb.(1)** | **Protein function(2)** | **Protein access. numbers(3)** | **Mass (Da)** | **pI** | | **Nb unique pept.** | **% seq. cov.** | **Peptide sequence** |
| (Figure 2a) | | | | | | | | |
| EB1 | Galectin-1 | P16045 | 14848 | 5.28 | S | 9 | 63% | DSNNLCLHFNPR - EDGTWGTEHR - FNAHGDANTIVCNTK - FNAHGDANTIVCNTKEDGTWGTEHR - LNMEAINYMAADGDFK - LNMEAINYMAADGDFKIK - LPDGHEFKFPNR - SFVLNLGK - VRGEVASDAK |
| EB2 | Coactosin-like protein | Q9CQI6 | 15926 | 5.23 | C | 10 | 49% | AAYNLVR - AGGANYDAQSE - ELEEDFIR - EVVQNFAK - FALITWIGEDVSGLQR - FTTGDAMSK - KAGGANYDAQSE - KELEEDFIR - LFAFVR - SKFALITWIGEDVSGLQR |
| EB3 | Eukaryotic translation initiation factor 5A-1 | P63242 | 16832 | 5.07 | N+C | - | C+N | MALDI/TOF PMF |
| EB4 | Protein mago nashi homolog | P61327 | 17146 | 5.74 | N+C | 2 | 21% | IGSLIDVNQSK - IIDDSEITKEDDALWPPPDR |
| EB5 | Prefoldin subunit 5 | Q9WU28 | 17338 | 5.94 | N | 7 | 53% | DCLNVLNK - ELLVPLTSSMYVPGK - IQPALQEK - IQQLTALGAAQATVK - KIDFLTK - NQLDQEVEFLSTSIAQLK - QAVMEMMSQK |
| EB6 | Ubiquitin-conjugating enzyme E2 N | P61089 | 17121 | 6.13 | N+C | 10 | 61% | DKWSPALQIR - ICLDILK - ICLDILKDK - IYHPNVDK - IYHPNVDKLGR - LELFLPEEYPMAAPK - LLAEPVPGIKAEPDESNAR - TNEAQAIETAR - WSPALQIR - YFHVVIAGPQDSPFEGGTFK |
| EB7 | Superoxide dismutase [Cu-Zn] | P08228 | 15924 | 6.02 | C | 6 | 46% | DGVANVSIEDR - GDGPVQGTIHFEQK - HVGDLGNVTAGK - QDDLGKGGNEESTK - TMVVHEKQDDLGK - VISLSGEHSIIGR |
| EB8 | Prefoldin subunit 2 | O70591 | 16516 | 6.2 | N+C+Mb | 3 | 23% | GAVSAEQVIAGFNR - IIETLSQQLQAK - MVGGVLVER |
| EB9 | Peptidyl-prolyl cis-trans isomerase A | P17742 | 17882 | 7.73 | C | 9 | 56% | EGMNIVEAMER - FEDENFILK - HTGPGILSMANAGPNTNGSQFFICTAK - IIPGFMCQGGDFTR - KITISDCGQL - SIYGEKFEDENFILK - VKEGMNIVEAMER - VSFELFADK - VSFELFADKVPK |
| EB10 | Nucleoside diphosphate kinase A | P15532 | 17190 | 6.84 | N+C | 10 | 76% | DRPFFTGLVK - EISLWFQPEELVEYK - FLQASEDLLK - GDFCIQVGR - GLVGEIIKR - NIIHGSDSVK - SAEKEISLWFQPEELVEYK - TFIAIKPDGVQR - VMLGETNPADSKPGTIR - YMHSGPVVAMVWEGLNVVK |
| EB11 | Stathmin | P54227 | 17257 | 5.75 | C | 4 | 31% | ASGQAFELILSPR - DLSLEEIQK - ESKDPADETEAD - ESVPDFPLSPPK |
| EB12 | Cofilin-1 | P18760 | 18542 | 8.22 | N+C+Mb | 4 | 37% | EILVGDVGQTVDDPYTTFVK - KEDLVFIFWAPENAPLK - LGGSAVISLEGKPL - YALYDATYETK |
| EB13 | Proteasome subunit beta type-9 | P28076 | 23379 | 5.07 | N+C | 2 | 10% | FTTNAITLAMNR - VSAGTAVVNR |
| EB14 | Transmembrane emp24 domain-containing protein 2 | Q9R0Q3 | 22172 | 5.08 | Mb | 2 | 11% | HEQEYMEVR - IVMFTIDIGEAPK |
| EB15 | ATP synthase subunit d | Q9DCX2 | 18732 | 5.52 | Mt | 10 | 64% | ANVAKPGLVDDFEK - IPVPEDK - IPVPEDKYTALVDQEEK - IQEYEK - LASLSEKPPAIDWAYYR - SCAEFVSGSQLR - SWNETFHAR - TIDWVSFVEVMPQNQK - YPYWPHQPIENL - YTALVDQEEK |
| EB16 | UMP-CMP kinase | Q9DBP5 | 22262 | 5.68 | N+C | 5 | 28% | EMDQTMAANAQK - FLIDGFPR - NPDSQYGELIEK - NQDNLQGWNK - YGYTHLSAGELLR |
| EB17 | Ferritin light chain 1 | P29391 | 20785 | 5.65 | C | 3 | 19% | LLEFQNDR - QNYSTEVEAAVNR - |
| | | | | | | | | TQEAMEAALAMEK |
| EB18 | Adenine phosphoribosyltransferase | P47957 | 19707 | 6.31 | C | 4 | 33% | GFLFGPSLAQELGVGCVLIR - IDYIAGLDSR - LPGPTVSASYSLEYGK - SFPDFPIPGVLFR |
| EB19 | Thioredoxin-dependent peroxide reductase | P20108 | 28109 | 7.15 | Mt | 6 | 28% | DYGVLLESAGIALR - GLFIIDPNGVVK - GTAVVNGEFK - HLSVNDLPVGR - NGGLGHMNITLLSDITK - SVEETLR |
| EB20 | NADH dehydrogenase [ubiquinone] iron-sulfur protein 8 | Q8K3J1 | 24021 | 5.89 | Mt | 8 | 38% | EPATINYPFEK - EQESEVDMK - FRGEHALR - GLGMTLSYLFR - ILMWTELIR - LCEAICPAQAITIEAEPR - YDIDMTK - YPSGEER |
| EB21 | Phosphatidylethanolamine-binding protein 1 | P70296 | 20812 | 5,19 | C | 3 | 28% | GNDISSGTVLSDYVGSGPPSGTGLHR - LYTLVLTDPDAPSR - VDYAGVTVDELGK |
| EB22 | Peroxiredoxin-2 | Q61171 | 21761 | 5.2 | C | 4 | 22% | EGGLGPLNIPLLADVTK - NDEGIAYR - QITVNDLPVGR - SVDEALR |
| EB23 | Translationally-controlled tumor protein | P63028 | 19445 | 4.76 | C | 3 | 23% | DLISHDELFSDIYK - EDGVTPFMIFFK - EIADGLCLEVEGK |
| EB24 | Rho GDP-dissociation inhibitor 2 | Q61599 | 22833 | 4.95 | C | 4 | 31% | DAQPQLEEADDDLDSK - ELQEMDKDDESLTK - TLLGDVPVVADPTVPNVTVTR - VNKDIVSGLK |
| EB25 | Rho GDP-dissociation inhibitor 1 | Q99PT1 | 23390 | 5.1 | C | 4 | 24% | AEEYEFLTPMEEAPK - TDYMVGSYGPR - VAVSADPNVPNVIVTR - YIQHTYR |
| EB26 | Tumor protein D52 | Q62393 | 24295 | 4.69 | RE | 3 | 17% | ASAAFSSVGSVITK - GWQDVTATNAYK - TSETLSQAGQK |
| EB27 | 14-3-3 protein beta/alpha | Q9CQV8 | 28069 | 4.77 | C | 6 | 24% | DSTLIMQLLR - NLLSVAYK - VISSIEQK - YDDMAAAMK - YLILNATQAESK - YLSEVASGENK |
| EB28 | Proteasome subunit alpha type-5 | Q9Z2U1 | 26393 | 4.74 | N+C | 2 | 11% | ITSPLMEPSSIEK - PFGVALLFGGVDEK |
| EB29 | 14-3-3 protein zeta/delta | P63101 | 27754 | 4.73 | C | 8 | 31% | DICNDVLSLLEK - FLIPNASQPESK - NLLSVAYK - SVTEQGAELSNEER - VVSSIEQK - YDDMAACMK - YLAEVAAGDDK - YLAEVAAGDDKK |
| EB30 | 14-3-3 protein gamma | P61982 | 28235 | 4.8 | C | 5 | 20% | NLLSVAYK - NVTELNEPLSNEER - VISSIEQK - YDDMAAAMK - YLAEVATGEK |
| EB31 | Elongation factor 1-beta | O70251 | 24676 | 4.53 | C | 4 | 24% | LAQYESK - SIQADGLVWGSSK - TPAGLQVLNDYLADK - YGPSSVEDTTGSGAADAK |
| EB32 | Tropomyosin alpha-3 chain | Q63610 | 28989 | 4.75 | C | 9 | 28% | EQAEAEVASLNR - EQAEAEVASLNRR - IQLVEEELDR - IQLVEEELDRAQER - KIQVLQQQADDAEER - KLVIIEGDLER - LATALQK - LVIIEGDLER - MELQEIQLK |
| EB33 | Ubiquitin carboxyl-terminal hydrolase isozyme L3 | Q9JKB1 | 26134 | 4.96 | C | 2 | 10% | FLENYDAIR - SQGQDVTSSVYFMK |
| EB34 | Inositol monophosphatase 1 | O55023 | 30419 | 5.08 | C | 4 | 17% | IIAANSITLAK - LQVSQQEDITK - SLLVTELGSSR - SSPADLVTVTDQK |
| EB35 | Chloride intracellular channel protein 1 | Q9Z1Q5 | 26996 | 5.09 | N+C+Mb | 4 | 27% | GVTFNVTTVDTK - LAALNPESNTSGLDIFAK - LFMVLWLK - VLDNYLTSPLPEEVDETSAEDEGISQR |
| EB36 | 6-phosphogluconolactonase | Q9CQ60 | 27237 | 5.55 | C | 2 | 13% | DLPAAAAPAGPASFAR - FALGLSGGSLVSMLAR |
| EB37 | Proteasome subunit alpha type-3 | O70435 | 28387 | 5,29 | N+C | 3 | 12% | AVENSSTAIGIR - SNFGYNIPLK - VFQVEYAMK |
| EB38 | NADH dehydrogenase | Q9DCT2 | 30131 | 6.67 | Mt | 3 | 13% | DFPLTGYVELR - FEIVYNLLSLR - VVAEPVELAQEFR |
| EB39 | Proteasome subunit alpha type-6 | Q9QUM9 | 27382 | 6.34 | N+C | 4 | 17% | AINQGGLTSVAVR - HITIFSPEGR - LYQVEYAFK - QTESTSFLEK |
| EB40 | Endoplasmic reticulum resident protein 29 | P57759 | 28807 | 5.9 | RE | 5 | 21% | ESYPVFYLFR - FDTQYPYGEK - GALPLDTVTFYK - ILDQGEDFPASEMAR - SLNILTAFR |
| EB41 | Superoxide dismutase [Mn] | P09671 | 24603 | 8.8 | Mt | 4 | 20% | AIWNVINWENVTER - GDVTTQVALQPALK - GELLEAIKR - NVRPDYLK |
| EB42 | Phosphoglycerate mutase 1 | Q9DBJ1 | 28787 | 6.67 | C | 8 | 36% | ALPFWNEEIVPQIK - AMEAVAAQGK - FSGWYDADLSPAGHEEAK - HGESAWNLENR - HYGGLTGLNK - KAMEAVAAQGK - VLIAAHGNSLR - YADLTEDQLPSCESLK |
| EB43 | GTP-binding nuclear protein Ran | P62827 | 24405 | 7.01 | N+C | 7 | 34% | FNVWDTAGQEK - HLTGEFEK - KYVATLGVEVHPLVFHTNR - LVLVGDGGTGK - NLQYYDISAK - SNYNFEKPFLWLAR - |
| | | | | | | | | YVATLGVEVHPLV FHTNR |
| EB44 | Triosephosphate isomerase | P17751 | 32173 | 5.56 | C | 5 | 20% | HVFGESDELIGQK - IIYGGSVTGATCK - SNVNDGVAQSTR - VTNGAFTGEISPG MIK - VVFEQTK |
| EB45 | S-formylglutathione hydrolase | Q9R0P3 | 31302 | 6.7 | C | 2 | 12% | AFSGYLGPDESK - SGYQQAASEHGLV VIAPDTSPR |
| EB46 | EF-hand domain-containing protein D2 | Q4FZY0 | 26902 | 5.01 | Mb | 2 | 9% | DGFIDLMELK - LSEIDVSTEGVK |
| EB47 | Microtubule-associated protein RP/EB family member 1 | Q61166 | 29999 | 5.12 | C+Golgi | 4 | 16% | AGPGMVR - EYDPVAAR - ILQAGFK - QGQETAVAPSLVA PALSKPK |
| EB48 | Annexin A5 | P48036 | 35736 | 4.82 | Mb | 2 | 7% | GTVTDFPGFDGR - VLTEIIASR |
| EB49 | Proliferating cell nuclear antigen | P17918 | 28731 | 4.66 | N | 2 | 9% | CAGNEDIITLR - NLAMGVNLTSMSK |
| EB50 | Eukaryotic translation initiation factor 3 subunit J | Q66JS6 | 29469 | 4.69 | C | 2 | 10% | ITNSLTVLCSEK - SLYYASFLEALVR |
| EB51 | Elongation factor 1-delta | P57776 | 31275 | 4.91 | C | 2 | 13% | ITSLEVENQNLR - SLAGSSGPGASSG PGGDHSELIVR |
| EB52 | Guanine nucleotide-binding protein G(i) subunit alpha-2 | P08752 | 40472 | 5.28 | C+Mb | 13 | 40% | AMGNLQIDFADPQ R - AVVYSNTIQSIMAI VK - EYQLNDSAAYYLN DLER - FEDLNK - FEDLNKR - IAQSDYIPTQQDVL R - ITQSSLTICFPEYT GANK - LFDSICNNK - LLLLGAGESGK - LWADHGVQACFG R - MFDVGGQR - SREYQLNDSAAYY LNDLER - YDEAASYIQSK |
| EB53 | Farnesyl pyrophosphate synthase | Q920E5 | 40565 | 5.48 | C | 2 | 7% | GLTVVQAFQELVE PK - LDAYNQEK |
| EB54 | Serine-threonine kinase receptor-associated protein | Q9Z1Z2 | 38421 | 4.99 | N+C | 3 | 12% | FSPDGELYASGSE DGTLR - GAVWGATLNK - YDYNSGEELESYK |
| EB55 | Galactokinase | Q9R0N0 | 42158 | 5.17 | C | 3 | 7% | LAVLITNSNVR - MEELEAGR - TAQAAAMSR |
| EB56 | 40S ribosomal protein SA | P14206 | 32866 | 4.8 | N+C+Mb | 6 | 29% | ADHQPLTEASYVN LPTIALCNTDSPLR - AIVAIENPADVSVIS SR - FTPGTFTNQIQAAF R - GAHSVGLMWWML AR - KSDGIYIINLK - SDGIYIINLK |
| EB57 | Tropomodulin-3 | Q9JHJ0 | 39487 | 5.02 | C | 3 | 10% | FGYQFTQQGPR - |
| | | | | | | | | LVEVNLNNIK - SNDPVAVAFADMLK |
| EB58 | Eukaryotic initiation factor 4A-I | P60843 | 46137 | 5.02 | C | 5 | 14% | ATQALVLAPTR - ELAQQIQK - LQMEAPHIIVGTPGR - MFVLDEADEMLSR - QFYINVER |
| EB59 | Dolichyl-diphosphooligosaccharide--protein glycosyltransferase | O54734 | 48775 | 5.52 | RE+Mb | 3 | 7% | APTIVGK - LPDVYGVFQFK - NTLLIAGLQAR |
| EB60 | Malate dehydrogenase | P14152 | 36494 | 6.16 | C | 5 | 18% | DLDVAVLVGSMPR - FVEGLPINDFSR - GEFITTVQQR - LGVTADDVK - VIVVGNPANTNCLTASK |
| EB61 | 26S proteasome non-ATPase regulatory subunit 14 | O35593 | 34559 | 6.06 | N+C | 11 | 58% | AGVPMEVMGLMLGEFVDDYTVR - AVAVVVDPIQSVK - EMLELAK - HYYSITINYR - LGGGMPGLGQGPPTDAPAVDTAEQVYISSLALLK - LINANMMVLGHEPR - MTPEQLAIK - QTTSNLGHLNKPSIQALIHGLNR - SWMEGLTLQDYSEHCK - VIDVFAMPQSGTGVSVEAVDPVFQAK - VVIDAFR |
| EB62 | Aldose reductase | P45376 | 35715 | 6.71 | C | 4 | 12% | EVGVALQEK - TIGVSNFNPLQIER - TTAQVLIR - VAIDLGYR |
| EB63 | Annexin A2 | P07356 | 38660 | 7.55 | Mb+sec | 5 | 17% | QDIAFAYQR - RAEDGSVIDYELIDQDAR - SLYYYIQQDTK - TNQELQEINR - TPAQYDASELK |
| EB64 | L-lactate dehydrogenase A chain | P06151 | 36481 | 7.61 | C | 7 | 18% | DYCVTANSK - LNLVQR - LVIITAGAR - QVVDSAYEVIK - SADTLWGIQK - VTLTPEEEAR - YLMGER |
| EB65 | Glyceraldehyde-3-phosphate dehydrogenase | P16858 | 35792 | 8.44 | N+C | 7 | 27% | GAAQNIIPASTGAAK - IVSNASCTTNCLAPLAK - LISWYDNEYGYSNR - PITIFQER - VKVGVNGFGR - VPTPNVSVVDLTCR - VVDLMAYMASK |
| EB66 | Phosphoglycerate kinase 1 | P09411 | 44545 | 8.02 | C | 3 | 9% | ALESPERPFLAILGGAK - VDFNVPMK - VLPGVDALSNV |
| EB67 | Poly(rC)-binding protein 1 | P60335 | 37480,2 | 6.66 | N | 2 | 8% | AITIAGVPQSVTEcVK-IITLTGPTNAIFK |
| EB68 | Transaldolase | Q93092 | 37370 | 6.57 | C | 7 | 21% | LGGPQEEQIK - LSFDKDAMVAR - LSSTWEGIQAGK - |
| | | | | | | | | MESALDQLK - SYEPQEDPGVK - TIVMGASFR - VSTEVDAR |
| EB69 | 26S protease regulatory subunit 8 | P62196 | 45609 | 7.11 | N+C | 3 | 9% | LEGGSGGDSEVQR - VPDSTYEMIGGLDK - VSGSELVQK |
| EB70 | Hematopoietic prostaglandin D synthase | Q9JHF7 | 23211 | 6.31 | C | 2 | 13% | MFNELLTHQAPR - VQAIPAISAWILK |
| EB71 | Glutamate dehydrogenase 1 | P26443 | 61320,3 | 8.05 | Mt | 4 | 6% | LVEDLK- mVEGFFDR- TAAYVNAIEK- YNLGLDLR |
| EB72 | Elongation factor 1-gamma | Q9D8N0 | 50043 | 6.31 | C | 4 | 11% | ALIAAQYSGAQVR - ILGLLDTHLK - LDPGSEETQTLVR - WFLTCINQPQFR |
| EB73 | Elongation factor Tu | Q8BFR5 | 18818 | 7.23 | Mt | 3 | 17% | AEAGDNLGALVR - DPELGVK - GTVVTGTLER |
| EB74 | Eukaryotic initiation factor 4A-III | Q91VC3 | 46824 | 6.3 | N+C | 18 | 47% | DELTLEGIK - DVIAQSQSGTGK - EANFTVSSMHGDMPQK - ETQALILAPTR - FMTDPIR - GFKEQIYDVYR - GIYAYGFEKPSAIQQR - GLDVPQVSLIINYDLPNNR - KGVAINFVK - KLDYGQHVVAGTPGR - LDYGQHVVAGTPGR - MLVLDEADEMLNK - QFFVAVER - RDELTLEGIK - VDWLTEK - VFDMIR - VLISTDVWAR - YLPPATQVVLISATLPHEILEMTNK |
| EB75 | Alpha-enolase | P17182 | 47124 | 6.37 | C+Mb | 8 | 24% | AAVPSGASTGIYEALELR - DATNVGDEGGFAPNILENK - GNPTVEVDLYTAK - GVSQAVEHINK - IGAEVYHNLK - LMIEMDGTENK - VNVVEQEK - YITPDQLADLYK |
| EB76 | V-type proton ATPase subunit H | Q8BVE3 | 55837,5 | 6.18 | Vacuole | 2 | 4% | VSIFFDYAK- YNALLAVQK |
| EB77 | T-complex protein 1 subunit beta | P80314 | 57459 | 5.97 | C | 2 | 4% | GATQQILDEAER - LAVEAVLR |
| EB78 | Cytochrome b-c1 complex subunit 1 | Q9CZ13 | 52834 | 5.81 | Mt | 16 | 39% | EHTAYLIK - FTGSEIR - HLSSVSR - IPLAEWESR - IQEVDAQMLR - LCTSATESEVTR - MVLAAAGGVEHQQLLDLAQK - NALVSHLDGTTPVCEDIGR - NNGAGYFLEHLAFK - RIPLAEWESR - |
| | | | | | | | | SGMFWLR - TDLTDYLNR - VASEQSSHATCTV GVWIDAGSR - VYEEDAVPGLTPC R - YETEKNNGAGYFL EHLAFK - YFYDQCPAVAGYG PIEQLPDYNR |
| EB79 | T-complex protein 1 subunit alpha | P11983 | 60432 | 5.82 | C | 22 | 54% | AFHNEAQVNPER - EQLAIAEFAR - FATEAAITILR - GANDFMCDEMER - IACLDFSLQK - ICDDELILIK - IHPTSVISGYR - IIGINGDYFANMVV DAVLAVK - LGVQVVITDPEK - MLVDDIGDVTITND GATILK - QAGVFEPTIVK - SLHDALCVVK - SLLVIPNTLAVNAA QDSTDLVAK - SQNVMAAASIANIV K - SSFGPVGLDK - SVVPGGGAVEAAL SIYLENYATSMGS R - TSASIILR - VLCELADLQDK - VLCELADLQDKEV GDGTTSVVIIAAEL LK - YFVEAGAMAVR - YINENLIINTDELGR - YPVNSVNILK |
| EB80 | Protein disulfide-isomerase A3 precursor | P27773 | 56643 | 5.88 | REL | - | 54% | MALDI/TOF PMF |
| EB81 | Transketolase | P40142 | 67588 | 7.23 | C | - | 39% | MALDI/TOF PMF |
| EB82 | Lamin-A/C | Q9DC21 | 74193 | 6.54 | N | - | 20% | MALDI/TOF PMF |
| EB83 | Plastin-3 | Q99K51 | 70322 | 5.54 | C | - | 14% | MALDI/TOF PMF |
| EB84 | Elongation factor 2 | P58252 | 95298 | 6.41 | C | 10 | 12% | AGIIASAR - EGALCEENMR - EGIPALDNFLDKL - ETVSEESNVLCLS K - GEGQLSAAER - GGGQIIPTAR - GVQYLNEIK - IMGPNYTPGK - QFAEMYVAK - VFDAIMNFR |
| EB85 | DNA replication licensing factor MCM7 | Q61881 | 81194 | 5.98 | N | 2 | 3% | GSSGVGLTAAVLR - SITVVLEGENTR |
| EB86 | von Willebrand factor A domain-containing protein 5A | Q99KC8 | 87087 | 6.15 | C | - | 18% | MALDI/TOF PMF |
| EB87 | Vinculin | Q64727 | 116644 | 5.77 | C | - | 23% | MALDI/TOF PMF |
| EB88 | Gelsolin precursor | P13020 | 85888 | 5,83 | C+Sec | 16 | 27% | MALDI/TOF PMF |
| EB89 | Heat shock cognate 71 kDa protein | Q3U9G0 | 70827 | 5.37 | C | 18 | 55% | MALDI/TOF PMF |
| EB90 | 60 kDa heat shock protein | P63038 | 60972 | 5.91 | Mt | 13 | 25% | ALMLQGVDLLADA VAVTMGPK - CEFQDAYVLLSEK - DDAMLLK - GANPVEIR - GVMLAVDAVIAELK - IGIEIIK - LSDGVAVLK - NAGVEGSLIVEK - TVIIEQSWGSPK - VGEVIVTK - VGGTSDVEVNEK - VGLQVVAVK - VTDALNATR |
| EB91 | Heat shock protein HSP 90-Beta | P11499 | 83142 | 4,97 | Mt | 11 | 17% | MALDI/TOF PMF |
| EB92 | Importin-5 | Q8BKC5 | 123,575.6 | 4.82 | N+C | 7 | 7.02% | MALDI/TOF PMF |
| EB93 | Vimentin | P20152 | 51832 | 5.05 | C | 5 | 16% | EMEENFALEAANY QDTIGR - ILLAELEQLK - KVESLQEEIAFLK - LLQDSVDFSLADAI NTEFK - MALDIEIATYR |
| EB94 | Ubiquitin thioesterase | Q7TQI3 | 31267 | 4.85 | C | 2 | 9% | IQQEIAVQNPLVSE R - LLTSGYLQR |
| EB95 | Peroxiredoxin-6 | O08709 | 24854 | 5.71 | C+L | 11 | 67% | DFTPVCTTELGR - DINAYNGETPTEK - DLAILLGMLDPVEK - FHDFLGDSWGILF SHPR - KGESVMVVPTLSE EEAK - LIALSIDSVEDHLA WSK - LPFPIIDDK - PGGLLLGDEAPNF EANTTIGR - PVATPVDWK - VVDSLQLTGTK - VVFIFGPDK |

(Figure 2b)

| | | | | | | | | |
|---|---|---|---|---|---|---|---|---|
| T1 | Alpha-enolase | P17182 | 47124 | 6,37 | C | 3 | 8% | IGAEVYHNLK - LMIEMDGTENK - VNQIGSVTESLQA CK |
| T2 | Alpha-enolase | P17182 | 47124 | 6,37 | C | 8 | 23% | GNPTVEVDLYTAK - GVSQAVEHINK - IEEELGSK - IGAEVYHNLK - LAMQEFMILPVGA SSFR - LAQSNGWGVMVS HR - LMIEMDGTENK - VNQIGSVTESLQA CK |
| T3 | Pyruvate kinase isozymes M1/M2 | P52480 | 57828 | 7,18 | N+C | 12 | 29% | FGVEQDVDMVFAS FIR - GADFLVTEVENGG SLGSK - GSGTAEVELK - GVNLPGAAVDLPA VSEK - ITLDNAYMEK - IYVDDGLISLQVK - LAPITSDPTEAAAV GAVEASFK - LDIDPITAR - MQHLIAR - |
| | | | | | | | | NTGIICTIGPASR - SGMNVAR - VNLAMDVGK |
| T4 | Actin | P60710 | 41805 | 5,29 | C | 12 | 35% | AGFAGDDAPR - AVFPSIVGR - AVFPSIVGRPR - DLTDYLMK - DSYVGDEAQSK - EITALAPSTMK - GYSFTTTAER - HQGVMVGMGQK - IWHHTFYNELR - QEYDESGPSIVHR - SYELPDGQVITIGNER - VAPEEHPVLLTEAPLNPK |
| T5 | 26S proteasome non-ATPase regulatory subunit 14 | O35593 | 34559 | 6,06 | N+C | 2 | 5% | HYYSITINYR - VVIDAFR |
| T6 | 26S proteasome non-ATPase regulatory subunit 7 | P26516 | 37008 | 6,29 | N+C | 6 | 22% | DTTVGTLSQR - PELAVQK - SVVALHNLINNK - TNDQMVVVYLASLIR - VVGVLLGSWQK - VVVHPLVLLSVVDHFNR |
| T6 | Tranldolase | Q93092 | 37371 | 6,57 | C | 6 | 18% | ALAGCDFLTISPK - LGGPQEEQIK - LIELYK - LSSTWEGIQAGK - MELDQLK - SYEPQEDPGVK |
| T7 | Alcohol dehydrogenase [NADP+] | Q9JII6 | 36569 | 6,90 | C | 5 | 19% | ALEVLVAK - ALGLSNFNSR - GLEVTAYSPLGSSDR - HPDEPVLLEEPVVLALAEK - SPAQILLR |
| T8 | Glyceraldehyde-3-phphate dehydrogenase | P16858 | 35792 | 8,44 | N+C | 7 | 31% | GAAQNIIPASTGAAK - IVSNASCTTNCLAPLAK - LISWYDNEYGYSNR - PITIFQER - VIHDNFGIVEGLMTTVHAITATQK - VPTPNVSVVDLTCR - VVDLMAYMASK |
| T9 | Biliverdin reductase A | Q9CY64 | 33507 | 6,53 | C | 8 | 30% | ELGSLDNVR - FGFPAFSGISR - FGVVVVGVGR - FTASPLEEEK - LLGQVEDLAAEK - MTVQLETQNK - QISLEDALR - SGSLEEVPNVGVNK |
| T10 | Chloride intracellular channel protein 1 | Q9Z1Q5 | 26996 | 5,09 | N+C | 4 | 29% | GVTFNVTTVDTK - LAALNPESNTSGLDIFAK - NSNPALNDNLEK - VLDNYLTSPLPEEVDETEDEGISQR |
| T11 | Rho GDP-dissociation inhibitor 2 | Q61599 | 22833 | 4,97 | C | 3 | 23% | DAQPQLEEADDDLDSK - LNYKPPPQK - TLLGDVPVVADPTVPNVTVTR |
| T12 | Myosin regulatory light chain 12B | Q3THE2 | 19777 | 4,71 | C | 5 | 30% | EAFNMIDQNR - ELLTTMGDR - FTDEEVDELYR - GNFNYIEFTR - LNGTDPEDVIR |
| T13 | ATP synthase subunit d | Q9DCX2 | 18732 | 5,52 | Or+Mb | 4 | 28% | ANVAKPGLVDDFEK - SCAEFVSGSQLR - SWNETFHAR - YTALVDQEEK |

(Figure 2c)

| | | | | | | | | |
|---|---|---|---|---|---|---|---|---|
| N1 | 60 kDa heat shock protein | P63038 | 60972 | 5,91 | Or+Mb | 18 | 35% | AAVEEGIVLGGGCALLR - ALMLQGVDLLADAVAVTMGPK - DDAMLLK - GANPVEIR - GVMLAVDAVIAELK - GVMLAVDAVIAELKK - GYISPYFINTSK - IGIEIIK - KGVITVK - LSDGVAVLK - LVQDVANNTNEEAGDGTTTATVLAR - NAGVEGSLIVEK - TLNDELEIIEGMK - TVIIEQSWGSPK - VGEVIVTK - VGGTSDVEVNEK - VGLQVVAVK - VTDALNATR |
| N2 | WD repeat-containing protein 1 | O88342 | 66388 | 6,11 | C | 14 | 31% | AHDGGIYAISWSPDSTHLLSGDK - DHLLSISLSGYINYLDK - DIAWTEDSKR - FATADGQIFIYDGK - GPVTDVAYSHDGAFLAVCDASK - IAVVGEGR - LATGSDDNCAAFFEGPPFK - SIQCLTVHR - VFASLPQVER - VINSVDIK - VYSILASTLKDEGK - YAPSGFYIASGDISGK - YEYQPFAGK - YTNLTLR |
| N3 | Coronin-1A | O89053 | 50971 | 6,05 | C | 13 | 35% | ADQCYEDVR - ATPEPSGTPSSDTVSR - CEPIAMTVPR - DAGPLLISLK - DGALICTSCR - EPVITLEGHTK - FMALICEASGGGAFLVLPLGK - ILTTGFSR - KCEPIAMTVPR - KSDLFQEDLYPPTAGPDPALTAEEWLGGR - LDRLEETVQAK - NLNAIVQK - VSQTTWDSGFCA |
| | | | | | | | | VNPK |
| N4 | Actin-related protein 3 | Q99JY9 | 47340 | 5,61 | C | 11 | 28% | DITYFIQQLLR - DREVGIPPEQSLETAK - DYEEIGPSICR - EFSIDVGYER - HGIVEDWDLMER - HNPVFGVMS - KDYEEIGPSICR - LPACVVDCGTGYTK - LSEELSGGR - NIVLSGGSTMFR - PIDVQVITHHMQR |
| N5 | Vimentin | P20152 | 53671 | 5,06 | C | 17 | 42% | DNLAEDIMR - EEAESTLQSFR - EMEENFALEAANYQDTIGR - ETNLESLPLVDTHSK - FADLSEAANR - ILLAELEQLK - ISLPLPTFSSLNLR - LGDLYEEEMR - LLQDSVDFSLADAINTEFK - LQDEIQNMK - LQEEMLQR - MALDIEIATYR - NLQEAEEWYK - QDVDNASLAR - QVDQLTNDK - QVQSLTCEVDALK - VELQELNDR |
| N6 | 40S ribomal protein | P14206 | 32866 | 4,80 | N+C | 10 | 43% | AIVAIENPADVSVISSR - DPEEIEKEEQAAAEK - FAAATGATPIAGR - FLAAGTHLGGTNLDFQMEQYIYK - FTPGTFTNQIQAAFR - GAHSVGLMWWMLAR - KSDGIYIINLK - LLVVTDPR - SDGIYIINLK - YVDIAIPCNNK |
| N7 | N-myc-interactor | O35309 | 35217 | 4,98 | C | 10 | 27% | CHSVAVSPCIER - CSLDQSFAAYFK - FQVHVDISK - KLEAELQSDAR - KNNGGGEVEVVK - LEAELQSDAR - NGGGEVESVDYDR - NGGGEVESVDYDRK - NNGGGEVEVVK - VITFVETGVVDK |
| N7 | 40S ribomal protein | P14206 | 32866 | 4,80 | N+C | 3 | 18% | AIVAIENPADVSVISSR - FAAATGATPIAGR - FLAAGTHLGGTNLDFQMEQYIYK |
| N8 | Serine-threonine kinase receptor-associated protein | Q9Z1Z2 | 38425 | 4,99 | N+C | 12 | 43% | AATAAADFTAK - ALWCSDDK - EFLVAGGEDFK - FSPDGELYASGSEDGTLR - GAVWGATLNK - IYDLNKPEAEPK - |
| | | | | | | | | LWDHATMTEVK - LWQTVVGK - SFEAPATINSLHPEK - SIAFHVSLEPIK - VWDAVSGDELMTLAHK - YDYNSGEELESYK |
| N9 | Nucleophosmin | Q61937 | 32542 | 4,62 | N+C | 6 | 29% | DELHIVEAEAMNYEGSPIK - GPSSVEDIK - MSVQPTVSLGGFEITPPVVLR - MTDQEAIQDLWQWR - TVSLGAGAKDELHIVEAEAMNYEGSPIK - VDNDENEHQLSLR |
| N10 | Proliferating cell nuclear antigen | P17918 | 28768 | 4,66 | N | 9 | 45% | AEDNADTLALVFEAPNQEK - ATPLSPTVTLSMDVPLVVEYK - CAGNEDIITLR - DLSHIGDAVVISCAK - FSGELGNGNIK - IADMGHLK - LIQGSILK - MPSGEFAR - NLAMGVNLTSMSK |
| N11 | 40S ribomal protein | P14206 | 32866 | 4,80 | N+C | 7 | 32% | AIVAIENPADVSVISSR - DPEEIEKEEQAAAEK - FAAATGATPIAGR - FTPGTFTNQIQAAFR - GAHSVGLMWWMLAR - LLVVTDPR - YVDIAIPCNNK |
| N12 | Actin | P60710 | 41805 | 5,29 | C | 7 | 21% | DLTDYLMK - DSYVGDEAQSK - EITALAPSTMK - GYSFTTTAER - LDLAGR - QEYDESGPSIVHR - VAPEEHPVLLTEAPLNPK |
| N12 | Transcriptional activator protein Pur-beta | O35295 | 33885 | 5,35 | N | 5 | 21% | FFFDVGCNK - FGGAFCR - GGGGGGGGGPGGFQPAPR - GGGGGGGGPGGEQETQELASK - LTLSMAVAAEFR |
| N13 | Isocitrate dehydrogenase [NAD] subunit alpha | Q9D6R2 | 39621 | 6,27 | Or+Mb | 11 | 33% | APIQWEER - CSDFTEEICR - DMANPTALLLVMMLR - HMGLFDHAAK - IAEFAFEYAR - IEAACFATIK - LITEEASKR - MSDGLFLQK - NVTAIQGPGGK - TPIAAGHPSMNLLLR - TPYTDVNIVTIR |
| N14 | 60S acidic ribomal protein | P14869 | 34353 | 5,91 | N+C | 9 | 32% | AGAIAPCEVTVPAQNTGLGPEK - CFIVGADNVGSK - DMLLANK - GHLENNPALEK - GNVGFVFTK - GTIEILSDVQLIK - |
| | | | | | | | | IIQLLDDYPK - SNYFLK - TSFFQALGITTK |
| N15 | Tranldolase | Q93092 | 37371 | 6,57 | C | 9 | 28% | ALAGCDFLTISPK - LFVLFGAEILK - LGGPQEEQIK - LSFDKDAMVAR - LSSTWEGIQAGK - MELDQLK - TIVMGASFR - VSTEVDAR - WLHNEDQMAVEK |
| N16 | Eukaryotic translation initiation factor 6 | O55135 | 26492 | 4,63 | N+C | 7 | 31% | ASFENNCEVGCFAK - DSLIDSLT - ETEEILADVLK - HGLLVPNNTTDQELQHIR - LNEAKPSTIATSMR - NSLPDSVQIR - PSTIATSMR |
| N17 | Proteasome subunit alpha type-5 | Q9Z2U1 | 26393 | 4,74 | N+C | 8 | 44% | EELEEVIKDI - GPQLFHMDPSGTFVQCDAR - GVNTFSPEGR - ITSPLMEPSSIEK - LFQVEYAIEAIK - LGSTAIGIQTSEGVCLAVEK - PFGVALLFGGVDEK - SSLIILK |
| N18 | Thioredoxin-dependent peroxide reductase | P20108 | 28109 | 7,15 | Or+Mb | 6 | 28% | DYGVLLEGIALR - ELSLDDFK - GLFIIDPNGVVK - GTAVVNGEFK - HLSVNDLPVGR - KNGGLGHMNITLLSDITK |
| N19 | Proteasome subunit alpha type-2 | P49722 | 25881 | 8,39 | N+C | 6 | 32% | AANGVVLATEK - GYSFSLTTFSPSGK - HIGLVYSGMGPDYR - RYNEDLELEDAIHTAILTLK - VASVMQEYTQSGGVR - YNEDLELEDAIHTAILTLK |

**Table 1: Identification and localization of spots highlighted in the two nuclear protein extracts: Total protein extract (Fig. 2a) Total nuclear protein extract (Fig. 2b) and NaCl protein extract (Fig. 2c)**. Three independent growth cultures of J774 cells have been performed. For each culture, total protein extract (5% of cells, Fig. 2a) and total nuclear protein extract (95% of cells) has been prepared (see materials and methods). NaCl nuclear protein extraction has been performed on the 90% of the total nuclear protein extract (see Materials and Methods) (Fig. 2c) and 10% has been put aside (Fig. 2b). The patterns of the different extracts have been compared after separation on 2-DE Gel. Mass spectrometry identification has been performed on spots systematically enriched in a nuclear extract compared to the other one. (1) Spot numbers circled on figure 2, (2) Protein function described by UniProtKB, (3) Accession number from UniProKB, (4) Protein localization

annotated in UniProtKB. When the localization is not annotated, the localization has been determined using the WoLF PSORT program (http://www.psort.org/) [43] Abbreviations: N = Nucleus, C = cytoplasm, Mt = Mitochondria, R = Ribosome, Mb = Membrane, G = Golgi, RE = Reticulum Endoplasmic, L = Lysosome, E = Endosome, V = Vacuole, NA = Nucleic Acid binding, U = Unknown, Or = organelles.

## Table 2: Part A (pH gradient 4-8)

| Protein Identification | | | | | Protein localization(4) | Mass spectrometry analysis | | | Delta 2D analysis | | | | | |
|---|---|---|---|---|---|---|---|---|---|---|---|---|---|---|
| **Spot Nb. (1)** | **Access.nb.(2)** | **Protein function(3)** | **Mass (Da)** | **pI** | | **% C(5)** | **Nb pep.(6)** | **Mascot Score** | **Mass spectrometry Analysis** | **J774 (7)** | **SD J774 (8)** | **XS52 (7)** | **SD XS52 (8)** | **XS52 / J774(9)** |
| 1 | P23198 | Chromobox protein homolog 3 (HP1g) | 20842 | 5,13 | N | 22% | 4 | 253 | nLC-MS/MS | 0,05207 | 36,0 | 0,02508 | 16,74311 | 0,48 |
| 2 | Q8K4M5 | COMM domain-containing protein 1 | 20983 | 7,03 | N+C | 13% | 3 | 225 | nLC-MS/MS | 0,02358 | 0,7 | 0,00704 | 39,4465 | 0,30 |
| 3 | Q80UW8 | DNA-directed RNA polymerases I, II, and III subunit RPABC1 | 24555 | 5,69 | N | 12% | 3 | 185 | nLC-MS/MS | 0,04323 | 14,2 | 0,01421 | 12,34891 | 0,33 |
| 4 | Q91XN7 | Tropomyosin alpha isoform | 28495 | 4,71 | C | 29% | 9 | 590 | nLC-MS/MS | 0,1156 | 39,1 | 0,03459 | 49,08996 | 0,30 |
| 5 | Q63610 | Tropomyosin alpha-3 chain | 28989 | 4,75 | C | 68% | 28/31 | 394 | MALDI-MS | 0,36754 | 36,3 | 0,10089 | 58,64702 | 0,27 |
| 6 | P14206 | 40S ribosomal protein SA | 32817 | 4,80 | R | 24% | 6 | 453 | nLC-MS/MS | 0,04939 | 16,0 | 0,01159 | 42,09206 | 0,23 |
| 7 | Q5M9K0<br>Q8BGT7<br>Q8R081 | H2-Ke6 protein<br>Survival of motor neuron-related-splicing factor 30<br>Heterogeneous nuclear ribonucleoprotein L | 26572<br>26737<br>60085 | 6,10<br>6,78<br>6,65 | Mt<br>N<br>N+C | 38%<br>21%<br>5% | 7<br>4<br>3 | 555<br>241<br>207 | nLC-MS/MS | 0,12013 | 24,3 | 0,04638 | 39,73895 | 0,39 |
| 8 | P03336 | Gag polyprotein | 60521 | 8,12 | Mb | 20% | 15/22 | 139 | MALDI-MS | 0,13565 | 10,6 | 0,01595 | 18,60904 | 0,12 |
| 9 | O88569 | Heterogeneous nuclear ribonucleoproteins A2/B1 | 37380 | 8,97 | N+C | 28% | 10/13 | 137 | MALDI-MS | 0,14731 | 25,4 | 0,02104 | 19,36445 | 0,14 |
| 11 | O88569 | Heterogeneous nuclear ribonucleoproteins A2/B1 | 37380 | 8,97 | N+C | 61% | 23/26 | 335 | MALDI-MS | 0,12645 | 25,3 | 0,0503 | 33,0473 | 0,40 |
| 12 | Q0VG47 | Heterogeneous nuclear ribonucleoprotein A3 | 37063 | 8,46 | N+C | 29% | 9 | 532 | nLC-MS/MS | 0,08082 | 23,3 | 0,01774 | 54,42051 | 0,22 |
| 13 | Q2HJC9<br>Q9R059 | Polyglutamine-binding protein 1<br>Four and a half LIM domains 3 | 30328<br>31773 | 5,93<br>5,80 | N<br>N | 20%<br>16% | 5<br>4 | 235<br>222 | nLC-MS/MS | 0,038 | 38,5 | 0,00722 | 38,2851 | 0,19 |
| 14 | Q9DB05 | Alpha-soluble NSF attachment protein | 33168 | 5,30 | Mb | 33% | 10 | 630 | nLC-MS/MS | 0,07845 | 32,8 | 0,03556 | 40,29004 | 0,45 |
| 15 | P14206 | 40S ribosomal protein SA | 32817 | 4,80 | R | 40% | 11/16 | 166 | MALDI-MS | 0,15487 | 33,1 | 0,04558 | 23,75385 | 0,29 |
| 16 | O35309 | N-myc-interactor | 35213 | 4,98 | C | 29% | 10/16 | 124 | MALDI-MS | 0,05785 | 16,6 | 0,02553 | 26,82393 | 0,44 |
| 17 | O35295 | Transcriptional activator protein Pur-beta | 33881 | 5,35 | N | 33% | 13/19 | 163 | MALDI-MS | 0,05116 | 29,1 | 0,00918 | 19,81047 | 0,18 |
| 18 | O35295 | Transcriptional activator protein Pur-beta | 33881 | 5,35 | N | 9% | 3 | 101 | nLC-MS/MS | 0,06792 | 36,1 | 0,00717 | 45,98016 | 0,11 |
| | Q3UFS4 | Coiled-coil domain-containing protein 75 | 31064 | 5,06 | NA | 3% | 1 | 75 | | | | | | |
| 19 | Q69ZQ2 | Pre-mRNA-splicing factor ISY1 homolog | 32969 | 5,15 | N | 18% | 4 | 277 | nLC-MS/MS | 0,04984 | 22,0 | 0,02482 | 10,71716 | 0,50 |
| 20 | O35295 | Transcriptional activator protein Pur-beta | 33881 | 5,35 | N | 16% | 6/25 | 79 | MALDI-MS | 0,04663 | 30,4 | 0,0209 | 5,68007 | 0,45 |
| 21 | Q5XJV3 | Eukaryotic translation initiation factor 3, subunit F | 37960 | 5,33 | C | 31% | 9 | 592 | nLC-MS/MS | 0,07937 | 39,5 | 0,02793 | 29,93987 | 0,35 |
| | O88544 | COP9 signalosome complex subunit 4 | 46256 | 5,57 | N+C | 17% | 6 | 324 | | | | | | |
| 22 | Q9WUK4 | Replication factor C subunit 2 | 38700 | 6,04 | N | 42% | 14 | 851 | nLC-MS/MS | 0,08617 | 15,4 | 0,03923 | 50,16719 | 0,46 |
| 23 | P10107 | Annexin A1 | 38710 | 6,97 | N+C | 30% | 9 | 641 | nLC-MS/MS | 0,03389 | 12,4 | 0,00899 | 57,56362 | 0,27 |
| 24 | Q99J62 | Replication factor C subunit 4 | 39842 | 6,29 | N | 10% | 4 | 222 | nLC-MS/MS | 0,06822 | 8,5 | 0,01486 | 33,21598 | 0,22 |
| 25 | Q4VBE8 | WD repeat-containing protein 18 | 47181 | 6,43 | N+C | 9% | 4 | 213 | nLC-MS/MS | 0,01891 | 6,9 | 0,00591 | 55,43778 | 0,31 |
| 26 | Q8VDW0 | ATP-dependent RNA helicase DDX39 | 49036 | 5,46 | N | 5% | 3 | 126 | nLC-MS/MS | 0,03819 | 3,1 | 0,01902 | 13,60159 | 0,50 |
| 27 | P29758 | Ornithine aminotransferase, mitochondrial | 48324 | 6,19 | Mt | 22% | 10/19 | 114 | MALDI-MS | 0,01807 | 33,6 | 0,00469 | 25,20771 | 0,26 |
| 28 | P20152 | Vimentin | 53655 | 5,06 | C | 56% | 31/31 | 451 | MALDI-MS | 0,09493 | 35,0 | 0,0201 | 11,48413 | 0,21 |
| 29 | P20152 | Vimentin | 53655 | 5,06 | C | 59% | 33/34 | 451 | MALDI-MS | 0,19493 | 46,4 | 0,05444 | 30,42035 | 0,28 |
| 30 | P20152 | Vimentin | 53655 | 5,06 | C | 45% | 24/24 | 357 | MALDI-MS | 0,06948 | 30,5 | 0,01708 | 30,0705 | 0,25 |
| 31 | P14211 | Calreticulin | 47965 | 4,33 | RE | 20% | 7 | 351 | nLC-MS/MS | 0,09759 | 17,4 | 0,03709 | 29,31098 | 0,38 |
| 32 | P20152 | Vimentin | 53655 | 5,06 | C | 47% | 25/25 | 373 | MALDI-MS | 0,03104 | 20,1 | 0,01492 | 27,19775 | 0,48 |
| 33 | P29416 | Beta-hexosaminidase subunit alpha | 60560 | 6,09 | L | 15% | 9/9 | 134 | MALDI-MS | 0,13776 | 7,6 | 0,0462 | 16,69143 | 0,34 |
| 34 | Q91YW3 | DnaJ homolog subfamily C member 3 | 57428 | 5,61 | RE | 21% | 11 | 694 | nLC-MS/MS | 0,03322 | 35,9 | 0,00933 | 27,7837 | 0,28 |
| | O89053 | Coronin-1A | 50957 | 6,05 | Mb + C | 5% | 2 | 109 | | | | | | |
| 35 | Q08943 | FACT complex subunit SSRP1 | 80810 | 6,33 | N | 16% | 12 | 755 | nLC-MS/MS | 0,16 | 24,2 | 0,06 | 34,9 | 0,39 |
| 36 | P60122 | RuvB-like 1 | 50182 | 6,02 | N | 18% | 8/19 | 71 | MALDI-MS | 0,09064 | 11,6 | 0,03075 | 11,54538 | 0,34 |
| 37 | O89053 | Coronin-1A | 50957 | 6,05 | Mb + C | 48% | 23/36 | 284 | MALDI-MS | 0,20549 | 15,1 | 0,09217 | 1,63744 | 0,45 |
| 38 | P80314 | T-complex protein 1 subunit beta | 57441 | 5,97 | C | 19% | 10/32 | 78 | MALDI-MS | 0,06834 | 10,2 | 0,02822 | 11,53644 | 0,41 |
| | O89053 | Coronin-1A | 50957 | 6,05 | Mb + C | 19% | 9/32 | 62 | | | | | | |
| 39 | Q3MHE2 | U4/U6 small nuclear ribonucleoprotein Prp4 | 58362 | 7,06 | N | 9% | 4 | 259 | nLC-MS/MS | 0,02762 | 18,5 | 0,00624 | 48,09453 | 0,23 |
| 40 | Q99J39 | Malonyl-CoA decarboxylase | 54701 | 9,13 | P+C+Mt | 2% | 1 | 72 | nLC-MS/MS | 0,04761 | 60,5 | 0,00475 | 29,40514 | 0,10 |
| 41 | P48678 | Lamin-A/C | 74193 | 6,54 | N | 32% | 22/40 | 170 | MALDI-MS | 0,03365 | 1,3 | 0,01303 | 8,38142 | 0,39 |
| | Q9EQP2 | EH domain-containing protein 4 | 61441 | 6,33 | Mb | 33% | 20/40 | 159 | | | | | | |
| 42 | Q9EQP2 | EH domain-containing protein 4 | 61441 | 6,33 | Mb | 33% | 19/29 | 183 | MALDI-MS | 0,11765 | 5,6 | 0,04435 | 17,89611 | 0,38 |
| | P48678 | Lamin-A/C | 74193 | 6,54 | N | 14% | 10/29 | 70 | | | | | | |
| 43 | O08599 | Syntaxin-binding protein 1 | 67526 | 6,49 | Mt | 8% | 4 | 281 | nLC-MS/MS | 0,02407 | 19,3 | 0,0093 | 13,27276 | 0,39 |
| | Q64324 | Syntaxin-binding protein 2 | 66315 | 6,28 | Mt | 3% | 2 | 110 | | | | | | |
| 44 | P26041 | Moesin | 67735 | 6,22 | Mb + C | 49% | 40/53 | 400 | MALDI-MS | 0,07345 | 9,5 | 0,0334 | 5,51394 | 0,45 |
| 45 | Q8BH57 | WD repeat-containing protein 48 | 75959 | 6,73 | L+C+N | 9% | 6 | 335 | nLC-MS/MS | 0,02961 | 15,5 | 0,0096 | 41,26338 | 0,32 |
| 46 | P49717 | DNA replication licensing factor MCM4 | 96676 | 6,77 | N | 7% | 6 | 298 | nLC-MS/MS | 0,10631 | 19,4 | 0,04324 | 26,6694 | 0,41 |
| | P39054 | Dynamin-2 | 98084 | 7,02 | Mb + C | 4% | 4 | 154 | | | | | | |
| 47 | Q8VE90 | Transducin (Beta)-like 3 | 88324 | 6,28 | N | 6% | 5 | 291 | nLC-MS/MS | 0,02406 | 21,7 | 0,00746 | 10,33492 | 0,31 |
| | P49717 | DNA replication licensing factor MCM4 | 96676 | 6,77 | N | 5% | 6 | 272 | | | | | | |
| 48 | P49717 | DNA replication licensing factor MCM4 | 96676 | 6,77 | N | 6% | 6 | 321 | nLC-MS/MS | 0,01713 | 13,9 | 0,00513 | 11,99744 | 0,30 |
| | Q9DBR0 | A-kinase anchor protein 8 | 76247 | 5,04 | N | 3% | 2 | 109 | | | | | | |
| 49 | P08228 | Superoxide dismutase [Cu-Zn] | 15933 | 6,02 | C | 24% | 3 | 221 | nLC-MS/MS | 0,01 | 26,3 | 0,02 | 26,1 | 3,40 |
| 50 | P63166 | Small ubiquitin-related modifier 1 | 11550 | 5,35 | N+C | 27% | 3 | 156 | nLC-MS/MS | 0,03984 | 35,4 | 0,09351 | 8,47351 | 2,35 |
| 51 | Q91WV0 | Protein Dr1 | 19419 | 4,69 | N | 16% | 3 | 112 | nLC-MS/MS | 0,01164 | 40,3 | 0,02581 | 35,18678 | 2,22 |
| 52 | Q63829 | COMM domain-containing protein 3 | 22023 | 5,36 | N+C | 22% | 4 | 190 | nLC-MS/MS | 0,01346 | 26,1 | 0,03107 | 28,08899 | 2,31 |
| 53 | Q9CPP0 | Nucleoplasmin-3 | 19011 | 4,71 | N | 20% | 2 | 173 | nLC-MS/MS | 0,02891 | 9,0 | 0,07269 | 11,80719 | 2,51 |
| 54 | Q9CQ80 | Vacuolar protein-sorting- | 20735 | 5,97 | Mb+N+C | 27% | 5 | 246 | nLC-MS/MS | 0,00402 | 79,5 | 0,02081 | 34,03679 | 5,17 |
| | | associated protein 25 | | | | | | | | | | | | |
| 55 | Q6P8I4 | PEST proteolytic signal-containing nuclear protein | 18951 | 6,86 | N | 27% | 4 | 196 | nLC-MS/MS | 0,01194 | 53,5 | 0,03927 | 13,39824 | 3,29 |
| 56 | P67871 | Casein kinase II subunit beta | 24926 | 5,33 | C | 16% | 5 | 272 | nLC-MS/MS | 0,02585 | 37,0 | 0,05521 | 7,27223 | 2,14 |
| 57 | Q9CQE8 | UPF0568 protein | 28135 | 6,40 | N+C | 42% | 13/15 | 210 | MALDI-MS | 0,04238 | 9,3 | 0,10782 | 7,47932 | 2,54 |
| 58(10) | P30681 | High mobility group protein B2 | 24147 | 6,88 | N | 46% | 13/25 | 152 | MALDI-MS | 0,02338 | 10,9 | 0,07612 | 10,18283 | 3,26 |
| | Q9CXW3 | Calcyclin-binding protein | 26494 | 7,63 | N+C | 22% | 6/25 | 69 | | | | | | |
| 59(10) | P30681 | High mobility group protein B2 | 24147 | 6,88 | N | 55% | 19/36 | 203 | MALDI-MS | | | | | |
| 60 | Q9WUK2 | Eukaryotic translation initiation factor 4H | 27324 | 6,67 | C | 61% | 19/43 | 205 | MALDI-MS | 0,01255 | 7,5 | 0,04759 | 21,74888 | 3,79 |
| 61 | P63158 | High mobility group protein B1 | 24878 | 5,62 | N | 43% | 13/34 | 127 | MALDI-MS | 0,00551 | 43,5 | 0,0221 | 15,99995 | 4,01 |
| 62 | P63158 | High mobility group protein B1 | 24878 | 5,62 | N | 39% | 12/19 | 158 | MALDI-MS | 0,01263 | 82,0 | 0,03618 | 32,35689 | 2,86 |
| 63 | P63158 | High mobility group protein B1 | 24878 | 5,62 | N | 19% | 4 | 189 | nLC-MS/MS | 0,01683 | 28,3 | 0,03565 | 15,00733 | 2,12 |
| | Q8K0H5 | Transcription initiation factor TFIID subunit 10 | 21827 | 6,13 | N | 11% | 1 | 153 | | | | | | |
| 64 | P97371 | Proteasome activator complex subunit 1 | 28655 | 5,73 | C | 16% | 3 | 230 | nLC-MS/MS | 0,00335 | 34,8 | 0,01897 | 31,67332 | 5,66 |
| 65 | Q61166 | Microtubule-associated protein RP/EB family member 1 | 29997 | 5,12 | C | 30% | 8 | 491 | nLC-MS/MS | 0,01248 | 44,1 | 0,03792 | 10,32363 | 3,04 |
| 66 | Q61166 | Microtubule-associated protein RP/EB family member 1 | 29997 | 5,12 | C | 58% | 16/30 | 206 | MALDI-MS | 0,0932 | 41,0 | 0,22936 | 7,28292 | 2,46 |
| 67 | Q61166 | Microtubule-associated protein RP/EB family member 1 | 29997 | 5,12 | C | 58% | 17/29 | 221 | MALDI-MS | 0,06342 | 30,5 | 0,14202 | 10,19719 | 2,24 |
| 68 | Q9CS42 | Ribose-phosphate pyrophosphokinase 2 | 34764 | 6,15 | N | 20% | 6 | 424 | nLC-MS/MS | 0,02325 | 49,1 | 0,04843 | 18,54855 | 2,08 |
| | Q99N96 | 39S ribosomal protein L1, mitochondrial | 34977 | 7,70 | R | 16% | 5 | 307 | | | | | | |
| 69 | Q9JIY5 | Serine protease HTRA2, mitochondrial | 49318 | 9,60 | Mt | 22% | 10/12 | 139 | MALDI-MS | 0,04221 | 30,4 | 0,08613 | 4,51488 | 2,04 |
| 70 | Q5U4D9 | THO complex subunit 6 homolog | 37291 | 6,67 | C | 20% | 8/16 | 105 | MALDI-MS | 0,01763 | 58,7 | 0,04418 | 5,64107 | 2,51 |
| 71 | Q1JQB2 | Mitotic checkpoint protein BUB3 | 36931 | 6,36 | N | 46% | 16/20 | 237 | MALDI-MS | 0,03127 | 4,8 | 0,06688 | 11,348 | 2,14 |
| 72 | P47754<br>O35639 | F-actin-capping protein subunit alpha-2<br>Annexin A3 | 32947<br>36348 | 5,57<br>5,33 | Mb + C<br>Mb | 25%<br>20% | 6<br>5 | 443<br>303 | nLC-MS/MS | 0,02714 | 38,9 | 0,07793 | 8,67243 | 2,87 |
| 73 | P62137 | Serine/threonine-protein phosphatase PP1-alpha catalytic subunit | 37516 | 5,94 | N+C | 33% | 11/17 | 140 | MALDI-MS | 0,00768 | 117,0 | 0,02958 | 40,26799 | 3,85 |
| 74 | P57776<br>Q61937 | Elongation factor 1-delta<br>Nucleophosmin | 31274<br>32540 | 4,91<br>4,62 | C<br>N | 44%<br>29% | 14/24<br>8/24 | 166<br>76 | MALDI-MS | 0,02763 | 31,6 | 0,05588 | 22,4982 | 2,02 |
| 75 | P57776<br>Q61937 | Elongation factor 1-delta<br>Nucleophosmin | 31274<br>32540 | 4,91<br>4,62 | C<br>N | 28%<br>19% | 7<br>5 | 517<br>282 | nLC-MS/MS | 0,00655 | 66,1 | 0,04034 | 11,9592 | 6,16 |
| 76 | Q61937 | Nucleophosmin | 32540 | 4,62 | N | 26% | 6 | 361 | nLC-MS/MS | 0,21204 | 41,7 | 0,60076 | 1,7938 | 2,83 |
| 77 | Q8BK64<br>Q9D6J3 | Activator of 90 kDa heat shock protein ATPase homolog 1<br>Coiled-coil domain-containing protein 94 | 38093<br>35966 | 5,41<br>5,84 | RE+C<br>N+C | 21%<br>11% | 8<br>2 | 376<br>112 | nLC-MS/MS | 0,00687 | 19,7 | 0,01775 | 27,50911 | 2,58 |
| 78 | Q9D0R9 | WD repeat-containing protein 89 | 42443 | 5,36 | N+C | 5% | 2 | 119 | nLC-MS/MS | 0,00546 | 21,6 | 0,01948 | 29,94214 | 3,57 |
| 79 | Q4FJP2 | nmi protein | 35326 | 5,08 | N | 32% | 10/10 | 161 | MALDI-MS | 0,00909 | 36,7 | 0,0364 | 5,85905 | 4,00 |
| 80 | O54984 | Arsenical pump-driving ATPase | 38797 | 4,81 | N+C+RE | 11% | 4 | 242 | nLC-MS/MS | 0,01755 | 45,7 | 0,06836 | 5,33909 | 3,90 |
| 81 | Q9CX97 | WD repeat-containing protein 55 | 42584 | 4,74 | N | 18% | 8/9 | 120 | MALDI-MS | 0,0088 | 65,5 | 0,03961 | 18,89798 | 4,50 |
| 82 | Q9CX97 | WD repeat-containing protein 55 | 42584 | 4,74 | N | 8% | 4 | 199 | nLC-MS/MS | 0,01877 | 25,0 | 0,05443 | 9,28551 | 2,90 |
| 83 | Q99JX3 | Golgi reassembly-stacking protein 2 | 47009 | 4,68 | G | 7% | 3 | 198 | nLC-MS/MS | 0,01342 | 50,5 | 0,05793 | 7,05597 | 4,32 |
| 84 | Q0VGB7<br>Q9CXG3 | Serine/threonine-protein phosphatase 4 regulatory subunit 2<br>Peptidyl-prolyl cis-trans isomerase-like 4 | 46450<br>57195 | 4,52<br>5,79 | N+C<br>N | 22%<br>5% | 7<br>2 | 503<br>93 | nLC-MS/MS | 0,00733 | 106,3 | 0,02267 | 23,4178 | 3,09 |
| 85 | Q9D8N0<br>P63037 | Elongation factor 1-gamma<br>DnaJ homolog subfamily A member 1 | 50029<br>44839 | 6,31<br>6,65 | C<br>Mb | 31%<br>13% | 13/22<br>7/22 | 151<br>49 | MALDI-MS | 0,0499 | 30,6 | 0,10844 | 8,05757 | 2,17 |
| 86 | P17182<br>P50580 | Alpha-enolase<br>Proliferation-associated protein 2G4 | 47111<br>43671 | 6,37<br>6,41 | Mb + C<br>N+C | 38%<br>12% | 14/21<br>5/21 | 177<br>57 | MALDI-MS | 0,01015 | 37,2 | 0,0426 | 6,17812 | 4,20 |
| 87 | Q61RT4 | Eukaryotic translation initiation factor 3, subunit F | 37857 | 5,33 | C | 27% | 9/15 | 124 | MALDI-MS | 0,03585 | 32,5 | 0,07936 | 12,47152 | 2,21 |
| | Q8CCS6 | Polyadenylate-binding protein 2 | 32277 | 5,13 | N+C | 16% | 4/15 | 59 | | | | | | |
| 88 | P97855 | Ras GTPase-activating protein-binding protein 1 | 51797 | 5,41 | Mb+N+C | 44% | 16/21 | 214 | MALDI-MS | 0,00799 | 47,3 | 0,02927 | 5,51581 | 3,66 |
| 89 | P97855 | Ras GTPase-activating protein-binding protein 1 | 51797 | 5,41 | Mb+N+C | 18% | 7/11 | 87 | MALDI-MS | 0,00312 | 39,3 | 0,02023 | 12,40136 | 6,49 |
| 90 | Q9WUA2 | Phenylalanyl-tRNA synthetase beta chain | 65628 | 6,69 | C | 13% | 8 | 501 | nLC-MS/MS | 0,03421 | 6,9 | 0,08774 | 7,38061 | 2,56 |
| 91 | Q9Z110 | Delta-1-pyrroline-5-carboxylate synthetase | 87242 | 7,18 | Mb+Mt | 14% | 11 | 718 | nLC-MS/MS | 0,01546 | 18,6 | 0,03119 | 11,49563 | 2,02 |
| 92 | Q8BIQ5 | Cleavage stimulation factor 64 kDa subunit | 61302 | 6,36 | N | 33% | 16/17 | 240 | MALDI-MS | 0,02326 | 12,2 | 0,04788 | 8,67393 | 2,06 |
| 93 | Q8BIQ5 | Cleavage stimulation factor 64 kDa subunit | 61302 | 6,36 | N | 38% | 21/30 | 261 | MALDI-MS | 0,00858 | 54,5 | 0,04178 | 11,48926 | 4,87 |
| 94 | Q6NVF9 | Cleavage and polyadenylation specificity factor subunit 6 | 59116 | 6,66 | N | 11% | 6 | 335 | nLC-MS/MS | 0,01812 | 42,4 | 0,04714 | 18,10188 | 2,60 |
| 95 | P11983 | T-complex protein 1 subunit alpha B | 60411 | 5,82 | C | 22% | 12 | 699 | nLC-MS/MS | 0,01426 | 25,8 | 0,02926 | 10,16248 | 2,05 |
| 96 | P97855 | Ras GTPase-activating protein-binding protein 1 | 51797 | 5,41 | N+C | 50% | 19/28 | 238 | MALDI-MS | 0,02333 | 38,7 | 0,05747 | 14,52695 | 2,46 |
| 97 | P61979 | Heterogeneous nuclear ribonucleoprotein K | 50944 | 5,39 | N+C | 30% | 14/18 | 177 | MALDI-MS | 0,07114 | 23,7 | 0,15447 | 4,49423 | 2,17 |
| 98 | Q9WVE8 | Protein kinase C and casein kinase substrate in neurons protein 2 | 55798 | 5,10 | C | 27% | 12 | 728 | nLC-MS/MS | 0,0137 | 44,3 | 0,05161 | 3,50428 | 3,77 |
| 99 | Q9WVE8 | Protein kinase C and casein kinase substrate in neurons protein 2 | 55798 | 5,10 | C | 17% | 9 | 569 | nLC-MS/MS | 0,01062 | 39,3 | 0,02632 | 26,51316 | 2,48 |
| 100 | Q60960 | Importin subunit alpha-1 | 60144 | 4,93 | N+C | 14% | 7 | 443 | nLC-MS/MS | 0,0084 | 35,3 | 0,0394 | 18,90786 | 4,69 |
| | Q9WVE8 | Protein kinase C and casein kinase substrate in neurons protein 2 | 55798 | 5,10 | C | 10% | 5 | 302 | | | | | | |
| 101 | Q8BMP6 | Golgi resident protein GCP60 | 60144 | 5,07 | G | 20% | 9 | 628 | nLC-MS/MS | 0,0194 | 38,3 | 0,04261 | 11,35742 | 2,20 |

## Part B (pH gradient 3.7-10.5)

| Protein Identification | | | | | Protein localization (4) | Mass spectrometry analysis | | | | Delta 2D analysis | | | | |
|---|---|---|---|---|---|---|---|---|---|---|---|---|---|---|
| **Spot nb. (1)** | **Access. nb. (2)** | **Protein function(3)** | **Mass (Da)** | **pI** | | **% C (5)** | **Nb pep. (6)** | **Mascot Score** | **Mass spec Analysis** | **J774 (7)** | **SD J774 (8)** | **XS52 (7)** | **SD XS52 (8)** | **XS52 / J774 (9)** |
| 102 | Q8BQ03 | Putative uncharacterized protein (DNA replication licensing factor MCM5) | 82296 | 8,65 | N | 11% | 9 | 483 | nLC-MS/MS | 0,03 | 58,0 | 0,01 | 70,2 | 0,35 |
| 103 | P49718 | DNA replication licensing factor MCM5 | 82290 | 8,70 | N | 8% | 6/7 | 61 | MALDI-MS | 0,03 | 54,3 | 0,01 | 81,2 | 0,20 |
| 104 | P49718 | DNA replication licensing factor MCM5 | 82290 | 8,70 | N | 16% | 15/20 | 129 | MALDI-MS | 0,05 | 73,3 | 0,01 | 47,5 | 0,21 |
| 105 | P10126 | Elongation factor 1-alpha 1 | 50082 | 9,10 | C | 5% | 3 | 158 | nLC-MS/MS | 0,09 | 14,8 | 0,04 | 63,5 | 0,41 |
| 106 | P09405 | Nucleolin | 76677 | 4,69 | N+C | 10% | 9/17 | 83 | MALDI-MS | 0,24 | 23,9 | 0,04 | 30,0 | 0,15 |
| 107 | P20152 | Vimentin | 53655 | 5,06 | C | 11% | 6 | 368 | nLC-MS/MS | 0,08 | 22,1 | 0,02 | 32,0 | 0,21 |
| 108 | P20152 | Vimentin | 53655 | 5,06 | C | 48% | 26 | | MALDI-MS | 0,12 | 29,0 | 0,02 | 69,9 | 0,13 |
| 109 | P20152 | Vimentin | 53655 | 5,06 | C | 52% | 28 | | MALDI-MS | 0,17 | 50,7 | 0,02 | 40,9 | 0,11 |
| 110 | O35309<br>Q9Z204 | N-myc-interactor<br>Heterogeneous nuclear ribonucleoproteins C1/C2 | 35213<br>34364 | 4,98<br>4,92 | C<br>N | 26%<br>14% | 8<br>4 | 398<br>267 | nLC-MS/MS | 0,17 | 8,5 | 0,04 | 28,4 | 0,24 |
| 111 | Q99L47 | Hsc70-interacting protein | 41629 | 5,19 | C | 4% | 2 | 80 | nLC-MS/MS | 0,03 | 58,8 | 0,00 | 70,2 | 0,13 |
| 112 | Q69ZQ2 | Pre-mRNA-splicing factor ISY1 homolog | 32969 | 5,15 | N | 26% | 7 | 337 | nLC-MS/MS | 0,04 | 33,7 | 0,02 | 22,4 | 0,46 |
| 113 | O35295 | Transcriptional activator protein Pur-beta | 33881 | 5,35 | N | 24% | 8/13 | 122 | MALDI-MS | 0,06 | 6,4 | 0,02 | 22,1 | 0,34 |
| 114 | Q9DCH4 | Eukaryotic translation initiation factor 3 subunit F | 37976 | 5,33 | C | 28% | 9 | 554 | nLC-MS/MS | 0,03 | 44,1 | 0,01 | 16,6 | 0,28 |
| 115 | Q9WUK4 | Replication factor C subunit 2 | 38700 | 6,04 | N | 38% | 12 | 768 | nLC-MS/MS | 0,07 | 12,9 | 0,03 | 27,2 | 0,43 |
| 116 | O88544 | COP9 signalosome complex subunit 4 | 46256 | 5,57 | N+C | 7% | 3 | 136 | nLC-MS/MS | 0,18 | 26,0 | 0,09 | 13,8 | 0,50 |
| 117 | Q99J62 | Replication factor C subunit 4 | 39842 | 6,29 | N | 20% | 7 | 412 | nLC-MS/MS | 0,03 | 25,3 | 0,01 | 43,1 | 0,24 |
| 118 | O88569 | Heterogeneous nuclear ribonucleoproteins A2/B1 | 37380 | 8,97 | N+C | 41% | 15 | | MALDI-MS | 0,11 | 21,4 | 0,02 | 12,0 | 0,19 |
| 119 | P68040 | Guanine nucleotide-binding protein subunit beta-2-like 1 | 35055 | 7,60 | N+C+Mb | 21% | 7 | | MALDI-MS | 0,10 | 22,4 | 0,02 | 33,3 | 0,17 |
| 120 | P48678 | Lamin-A/C | 74193 | 6,54 | N | 11% | 7 | 447 | nLC-MS/MS | 0,04 | 19,8 | 0,01 | 21,0 | 0,35 |
| 121 | O88569 | Heterogeneous nuclear ribonucleoproteins A2/B1 | 37380 | 8,97 | N+C | 16% | 6/8 | 80 | MALDI-MS | 0,02 | 53,1 | 0,00 | 121,4 | 0,02 |
| 122 | Q9D1J3 | Nuclear protein Hcc-1 | 23518 | 6,29 | N | 20% | 5 | 261 | nLC-MS/MS | 0,05 | 9,7 | 0,02 | 12,2 | 0,38 |
| 123 | P46737 | Lys-63-specific deubiquitinase BRCC36 | 33319 | 5,54 | N | 9% | 3 | 171 | nLC-MS/MS | 0,03 | 22,1 | 0,01 | 19,2 | 0,44 |
| 124 | P14206 | 40S ribosomal protein SA | 32817 | 4,80 | R | 33% | 9 | 571 | nLC-MS/MS | 0,07 | 36,0 | 0,04 | 15,9 | 0,49 |
| 125 | Q9DB05 | Alpha-soluble NSF attachment protein | 33168 | 5,30 | Mb | 12% | 3 | 220 | MALDI-MS | 0,07 | 35,0 | 0,03 | 32,6 | 0,46 |
| 126 | P14206 | 40S ribosomal protein SA | 32817 | 4,80 | R | 30% | 8 | | MALDI-MS | 0,18 | 32,6 | 0,06 | 25,8 | 0,33 |
| 127 | P23198 | Chromobox protein homolog 3 (HP1g) | 20842 | 5,13 | N | 15% | 2 | 166 | nLC-MS/MS | 0,14 | 6,4 | 0,07 | 14,7 | 0,50 |
| 128 | Q9Y5S9 | RNA-binding protein 8A | 19877 | 5,50 | N+C | 20% | 3 | 153 | nLC-MS/MS | 0,16 | 17,4 | 0,06 | 26,7 | 0,34 |
| 129 | Q9CWZ3 | RNA-binding protein 8A | 19877 | 5,50 | N+C | 21% | 3 | 187 | nLC-MS/MS | 0,10 | 17,3 | 0,02 | 34,0 | 0,17 |
| 130 | Q7TMY4 | THO complex subunit 7 homolog | 23700 | 5,46 | N+C | 26% | 6 | | MALDI-MS | 0,15 | 14,6 | 0,05 | 44,8 | 0,34 |
| 131 | P20108 | Thioredoxin-dependent peroxide reductase, mitochondrial | 28109 | 7,15 | Mt | 8% | 2 | 101 | nLC-MS/MS | 0,06 | 15,5 | 0,02 | 45,7 | 0,33 |
| 132 | P17742 | Peptidyl-prolyl cis-trans isomerase A | 17960 | 7,74 | C | 17% | 3 | 170 | nLC-MS/MS | 0,08 | 36,4 | 0,17 | 32,2 | 2,19 |
| 133 | Q8BG13 | Putative uncharacterized protein (Putative RNA-binding protein 3) | 16751 | 7,98 | N+C | 40% | 8/14 | 116 | MALDI-MS | 0,09 | 17,2 | 0,18 | 14,3 | 2,05 |
| 134 | P17742 | Peptidyl-prolyl cis-trans isomerase A | 17960 | 7,74 | C | 21% | 4 | 196 | nLC-MS/MS | 0,07 | 17,3 | 0,16 | 19,4 | 2,21 |
| 135 | P63166 | Small ubiquitin-related modifier 1 | 11550 | 5,35 | N+C | 41% | 6/8 | 88 | MALDI-MS | 0,03 | 21,4 | 0,09 | 22,5 | 2,97 |
| 136 | P63166 | Small ubiquitin-related modifier 1 | 11550 | 5,35 | N+C | 49% | 7/28 | 67 | MALDI-MS | 0,06 | 34,9 | 0,16 | 24,2 | 2,56 |
| 137 | Q61686 | Chromobox protein homolog 5 (HP1a) | 22172 | 5,71 | N | 16% | 3 | 190 | nLC-MS/MS | 0,03 | 6,1 | 0,07 | 15,3 | 2,11 |
| 138 | P67871 | Casein kinase II subunit beta | 24926 | 5,33 | N+C | 11% | 4 | 201 | nLC-MS/MS | 0,01 | 48,1 | 0,04 | 21,5 | 3,49 |
| 139 | P97372 | Proteasome activator complex subunit 2 | 27040 | 5,54 | N+C | 31% | 11 | | MALDI-MS | 0,04 | 25,2 | 0,08 | 10,6 | 2,02 |
| 140 | Q9R1P4<br>Q9CX56<br>Q8BTW3 | Proteasome subunit alpha type-1<br>26S proteasome non-ATPase regulatory subunit 8<br>Exosome complex exonuclease MTR3 | 29528<br>30007<br>28353 | 6,00<br>6,03<br>5,87 | N+C<br>N+C<br>N+C | 22%<br>9%<br>11% | 6<br>3<br>2 | 301<br>138<br>123 | nLC-MS/MS | 0,77 | 8,6 | 0,17 | 11,7 | 2,18 |
| 141 | Q9Z1Q5 | Chloride intracellular channel protein 1 | 26996 | 5,09 | N+C | 24% | 6 | 337 | nLC-MS/MS | 0,10 | 64,3 | 0,26 | 15,8 | 2,52 |
| 142 | Q61166 | Microtubule-associated protein RP/EB family member 1 | 29997 | 5,12 | C | 58% | 16 | | MALDI-MS | 0,10 | 35,8 | 0,28 | 3,5 | 2,77 |
| 143 | Q61166 | Microtubule-associated protein RP/EB family member 1 | 29997 | 5,12 | C | 55% | 15/24 | 193 | MALDI-MS | 0,08 | 27,3 | 0,18 | 6,6 | 2,29 |
| 144 | Q61166 | Microtubule-associated protein RP/EB family member 1 | 29997 | 5,12 | C | 29% | 8/11 | 117 | MALDI-MS | 0,04 | 44,0 | 0,11 | 8,6 | 2,54 |
| 145 | Q61937 | Nucleophosmin | 32540 | 4,62 | N | 43% | 12/18 | 174 | MALDI-MS | 0,04 | 61,4 | 0,28 | 5,7 | 7,51 |
| 146 | Q8BFQ4<br>P68040 | WD repeat-containing protein 82<br>Guanine nucleotide-binding protein subunit beta-2-like 1 | 35056<br>35055 | 7,59<br>7,60 | Mb+C<br>R | 6%<br>6% | 3<br>2 | 118<br>110 | nLC-MS/MS | 0,01 | 77,9 | 0,03 | 49,6 | 2,41 |
| 147 | P60335 | Poly(rC)-binding protein 1 | 37474 | 6,66 | N+C | 18% | 6/6 | 101 | MALDI-MS | 0,05 | 63,4 | 0,10 | 12,0 | 2,10 |
| 148 | Q60737 | Casein kinase II subunit alpha | 45133 | 7,79 | N+C | 8% | 4 | 204 | nLC-MS/MS | 0,03 | 6,4 | 0,09 | 4,3 | 2,65 |
| 149 | Q8BG05 | Heterogeneous nuclear ribonucleoprotein A3 | 39628 | 9,10 | N+C | 29% | 11 | | MALDI-MS | 0,09 | 40,9 | 0,20 | 15,5 | 2,16 |
| 150 | Q8QZT1 | Acetyl-CoA acetyltransferase | 44787 | 8,71 | Mt | 29% | 13 | | MALDI-MS | 0,03 | 23,9 | 0,10 | 22,7 | 2,94 |
| 151 | O55131<br>P10126 | Septin-7<br>Elongation factor 1-alpha 1 | 50518<br>50082 | 8,73<br>9,10 | C<br>C | 17%<br>13% | 9<br>7 | | MALDI-MS | 0,08 | 45,5 | 0,16 | 16,5 | 2,12 |
| 152 | Q9CY58 | Plasminogen activator inhibitor 1 RNA-binding protein | 44687 | 8,60 | N+C | 55% | 24 | | MALDI-MS | 0,07 | 15,2 | 0,14 | 4,9 | 2,03 |
| 153 | Q9CY58 | Plasminogen activator inhibitor 1 RNA-binding protein | 44687 | 8,60 | N+C | 35% | 19/26 | 211 | MALDI-MS | 0,04 | 26,2 | 0,08 | 16,7 | 2,22 |
| 154 | Q9CY58 | Plasminogen activator inhibitor 1 RNA-binding protein | 44687 | 8,60 | N+C | 49% | 24/30 | 280 | MALDI-MS | 0,05 | 70,0 | 0,13 | 17,7 | 2,77 |
| 155 | Q99K48<br><br>Q5RJV5 | Non-POU domain-containing octamer-binding protein<br>Polypyrimidine tract binding protein 1 | 54506<br><br>59227 | 9,01<br><br>9,28 | N<br><br>N | 19%<br><br>14% | 11/28<br><br>8/28 | 49<br><br>68 | MALDI-MS | 0,04 | 30,4 | 0,09 | 17,9 | 2,25 |
| 156 | Q99K48 | Non-POU domain-containing octamer-binding protein | 54506 | 9,01 | N | 60% | 40/50 | 348 | MALDI-MS | 0,03 | 25,0 | 0,11 | 23,2 | 3,77 |
| 157 | Q9D0E1 | Heterogeneous nuclear ribonucleoprotein M | 77597 | 8,80 | N | 59% | 58 | | MALDI-MS | 0,02 | 10,6 | 0,07 | 20,9 | 3,90 |
| 158 | Q9D0E1 | Heterogeneous nuclear ribonucleoprotein M | 77597 | 8,80 | N | 54% | 52 | | MALDI-MS | 0,03 | 31,5 | 0,09 | 22,7 | 2,75 |
| 159 | Q9D0E1 | Heterogeneous nuclear ribonucleoprotein M | 77597 | 8,80 | N | 30% | 22 | 1281 | nLC-MS/MS | 0,03 | 43,3 | 0,07 | 20,4 | 2,82 |
| 160 | Q9CW46 | Ribonucleoprotein PTB-binding 1 | 79333 | 8,91 | N+C | 14% | 12 | | MALDI-MS | 0,01 | 73,5 | 0,02 | 32,2 | 2,96 |
| 161 | Q9CW46 | Ribonucleoprotein PTB-binding 1 | 79333 | 8,91 | N+C | 16% | 13 | | MALDI-MS | 0,00 | 67,5 | 0,05 | 40,8 | 9,78 |
| 162 | Q9CW46 | Ribonucleoprotein PTB-binding 1 | 79333 | 8,91 | N+C | 32% | 21 | | MALDI-MS | 0,00 | 133,5 | 0,06 | 47,7 | 14,53 |
| 163 | Q9CW46 | Ribonucleoprotein PTB-binding 1 | 79333 | 8,91 | N+C | 13% | 11 | | MALDI-MS | 0,01 | 38,5 | 0,04 | 27,4 | 3,77 |
| 164 | Q501J6 | Probable ATP-dependent RNA helicase DDX17 | 72354 | 8,82 | N | 20% | 13/17 | 134 | MALDI-MS | 0,02 | 12,7 | 0,06 | 33,1 | 3,03 |
| 165 | Q61990 | Poly(rC)-binding protein 2 | 38197 | 6,33 | N | 13% | 4 | 246 | nLC-MS/MS | 0,01 | 31,5 | 0,02 | 4,9 | 2,14 |
| 166 | Q9Z1D1 | Eukaryotic translation initiation factor 3 subunit G | 35616 | 5,69 | N+C | 19% | 9/10 | 110 | MALDI-MS | 0,01 | 11,3 | 0,03 | 18,9 | 2,13 |
| 167 | P17182 | Alpha-enolase | 47111 | 6,37 | Mb+C | 18% | 8/18 | 86 | MALDI-MS | 0,01 | 26,0 | 0,03 | 5,2 | 3,12 |
| 168 | Q922R8<br><br>Q60973 | Protein disulfide-isomerase A6<br>Histone-binding protein RBBP7 | 48070<br><br>47760 | 5,00<br><br>4,89 | Mb+RE<br><br>N | 16%<br><br>13% | 6/12<br><br>6/12 | 68<br><br>67 | MALDI-MS | 0,03 | 36,2 | 0,09 | 8,4 | 3,24 |
| 169 | Q922R8 | Protein disulfide-isomerase A6 | 48070 | 5,00 | Mb+RE | 17% | 7 | | MALDI-MS | 0,03 | 21,1 | 0,11 | 6,3 | 3,32 |
| 170 | Q99L47<br><br>Q922R8 | Hsc70-interacting protein<br>Protein disulfide-isomerase A6 | 41629<br><br>48070 | 5,19<br><br>5,00 | C<br><br>Mb+RE | 38%<br><br>30% | 16<br><br>12 | | MALDI-MS | 0,10 | 14,8 | 0,21 | 3,6 | 2,03 |
| 171 | P54775 | 26S protease regulatory subunit 6B | 47252 | 5,18 | N+C | 8% | 4 | 226 | nLC-MS/MS | 0,02 | 38,2 | 0,05 | 13,0 | 2,72 |
| 172 | P54775 | 26S protease regulatory subunit 6B | 47252 | 5,18 | N+C | 20% | 10 | | MALDI-MS | 0,03 | 14,9 | 0,07 | 5,1 | 2,10 |
| | Q9Z2X1 | Heterogeneous nuclear ribonucleoprotein F | 45701 | 5,31 | N | 11% | 5 | | | | | | | |
| 173 | P61979 | Heterogeneous nuclear ribonucleoprotein K | 50944 | 5,39 | N | 24% | 11 | | MALDI-MS | 0,01 | 55,8 | 0,07 | 7,9 | 6,00 |
| 174 | P61979 | Heterogeneous nuclear ribonucleoprotein K | 50944 | 5,39 | N | 33% | 16 | | MALDI-MS | 0,05 | 11,8 | 0,16 | 9,5 | 3,00 |
| 175 | Q61233 | Plastin-2 | 70105 | 5,20 | C | 10% | 6 | 367 | nLC-MS/MS | 0,01 | 27,6 | 0,03 | 20,3 | 3,62 |
| 176 | Q61233 | Plastin-2 | 70105 | 5,20 | C | 13% | 8 | | MALDI-MS | 0,02 | 20,6 | 0,04 | 20,5 | 2,18 |
| 177 | Q61233 | Plastin-2 | 70105 | 5,20 | C | 13% | 8 | | MALDI-MS | 0,01 | 31,5 | 0,03 | 21,3 | 3,47 |
| 178 | P61979 | Heterogeneous nuclear ribonucleoprotein K | 50944 | 5,39 | N+C | 26% | 12 | | MALDI-MS | 0,02 | 29,6 | 0,07 | 4,2 | 3,65 |
| | Q63850 | Nuclear pore glycoprotein p62 | 53222 | 5,21 | N | 19% | 10 | | | | | | | |
| 179 | P97855 | Ras GTPase-activating protein-binding protein 1 | 51797 | 5,41 | N+C | 9% | 4 | 211 | nLC-MS/MS | 0,00 | 173,2 | 0,01 | 33,8 | 12,36 |
| | P61979 | Heterogeneous nuclear ribonucleoprotein K | 50944 | 5,39 | N | 7% | 3 | 199 | | | | | | |
| 180 | P97855 | Ras GTPase-activating protein-binding protein 1 | 51797 | 5,41 | N+C | 32% | 13 | | MALDI-MS | 0,01 | 32,4 | 0,04 | 6,1 | 3,45 |
| 181 | P97855 | Ras GTPase-activating protein-binding protein 1 | 51797 | 5,41 | N+C | 47% | 20 | | MALDI-MS | 0,01 | 45,6 | 0,03 | 12,9 | 4,74 |
| 182 | Q6NVF9 | Cleavage and polyadenylation specificity factor subunit 6 | 59116 | 6,66 | N | 18% | 11 | | MALDI-MS | 0,02 | 31,1 | 0,04 | 6,9 | 2,08 |
| 183 | Q6NVF9 | Cleavage and polyadenylation specificity factor subunit 6 | 59116 | 6,66 | N | 15% | 9 | | MALDI-MS | 0,02 | 36,0 | 0,04 | 10,2 | 2,48 |
| 184 | Q6NVF9 | Cleavage and polyadenylation specificity factor subunit 6 | 59116 | 6,66 | N | 16% | 10 | | MALDI-MS | 0,01 | 29,5 | 0,03 | 15,5 | 3,09 |
| 185 | Q8BIQ5 | Cleavage stimulation factor 64 kDa subunit | 61302 | 6,36 | N | 17% | 11 | | MALDI-MS | 0,00 | 55,1 | 0,04 | 7,5 | 10,27 |
| 186 | Q8BIQ5 | Cleavage stimulation factor 64 kDa subunit | 61302 | 6,36 | N | 18% | 13 | | MALDI-MS | 0,02 | 59,8 | 0,04 | 8,8 | 2,72 |
| 189 | Q8BIQ5 | Cleavage stimulation factor 64 kDa subunit | 61302 | 6,36 | N | 12% | 8 | | MALDI-MS | 0,02 | 28,7 | 0,04 | 14,9 | 2,25 |
| 190 | Q8CCF0 | U4/U6 small nuclear ribonucleoprotein Prp31 | 55367 | 5,55 | N | 13% | 6 | | MALDI-MS | 0,04 | 10,5 | 0,07 | 9,0 | 2,02 |
| 191 | Q91WJ8 | Far upstream element-binding protein 1 | 68497 | 7,74 | N | 8% | 5 | 267 | nLC-MS/MS | 0,00 | 106,9 | 0,01 | 58,6 | 21,26 |
| 192 | P52480 | Pyruvate kinase isozymes M1/M2 | 57808 | 7,18 | N+C | 11% | 8 | | MALDI-MS | 0,03 | 53,5 | 0,07 | 7,2 | 2,27 |
| 193 | P50580 | Proliferation-associated protein 2G4 | 43671 | 6,41 | N+C | 10% | 4 | 241 | nLC-MS/MS | 0,00 | 62,9 | 0,02 | 24,5 | 4,13 |

**Table 2: Characterization of spots highlighted on the comparison of NaCl nuclear protein extracts from J774 and XS52 cell lines (Fig. 4)**

(1) Spot number circled on the Figure 4a (Part A) and Figure 4b (Part B), (2) Accession number from UniProKB, (3) Protein function described by UniProtKB, (4) Protein localization annotated in UniProtKB. When the localization is not annotated, the localization has been determined using the WoLF PSORT program (http://www.psort.org/) [43] Abbreviations: N = Nucleus, C = cytoplasm, Mt = Mitochondria, R = Ribosome, Mb = Membrane, G = Golgi, RE = Reticulum Endoplasmic, L = Lysosome, E = Endosome, V = Vacuole, NA = Nucleic Acid binding, U = Unknown, (5) Percentage of coverage, (6) When two numbers are noted, the first number indicates the number of matched peaks and the second, the number of unmatched peaks, (7) average quantification of the spot using Delta2D software from three independent 2-DE Gels in a cell line, (8) Standard Deviation of the considering spot in a cell line and (9) Ratio of the average quantification determined by the Delta 2D analysis: XS52/J774. (10) Spots 80 and 81 have been detected together by the delta 2D software

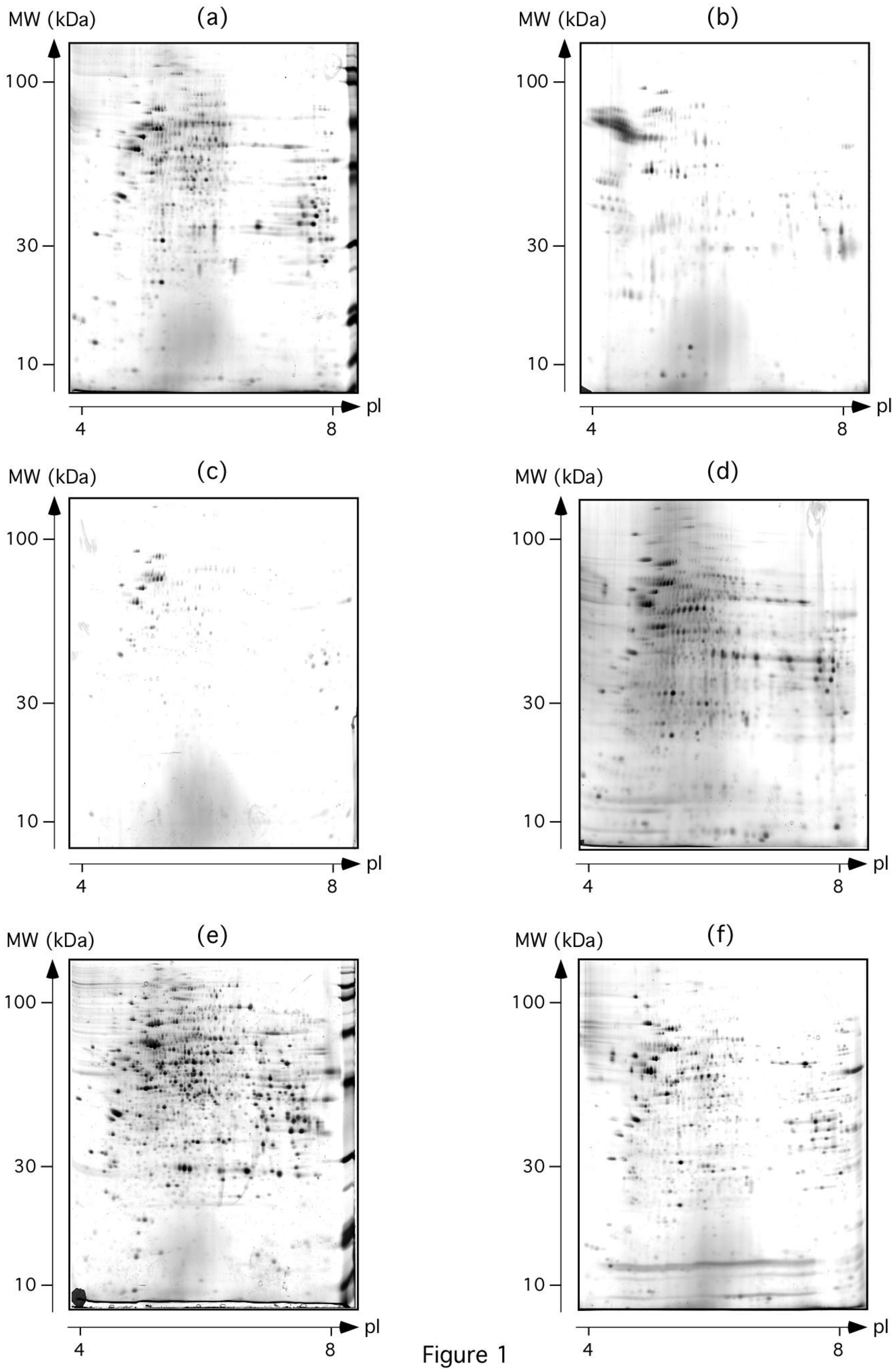

**Figure 1: Test of different nuclear protein extract preparations**

J774 cells were disrupted and nuclei were prepared as described by Rabilloud and coll. [6] with some modifications (see Materials & Methods). Then, nuclear proteins were extracted using (a) NaCl/SB3-12, (b) DNase, (c) Urea, (d) Benzonase, (e) NaCl and (f) lecithin.

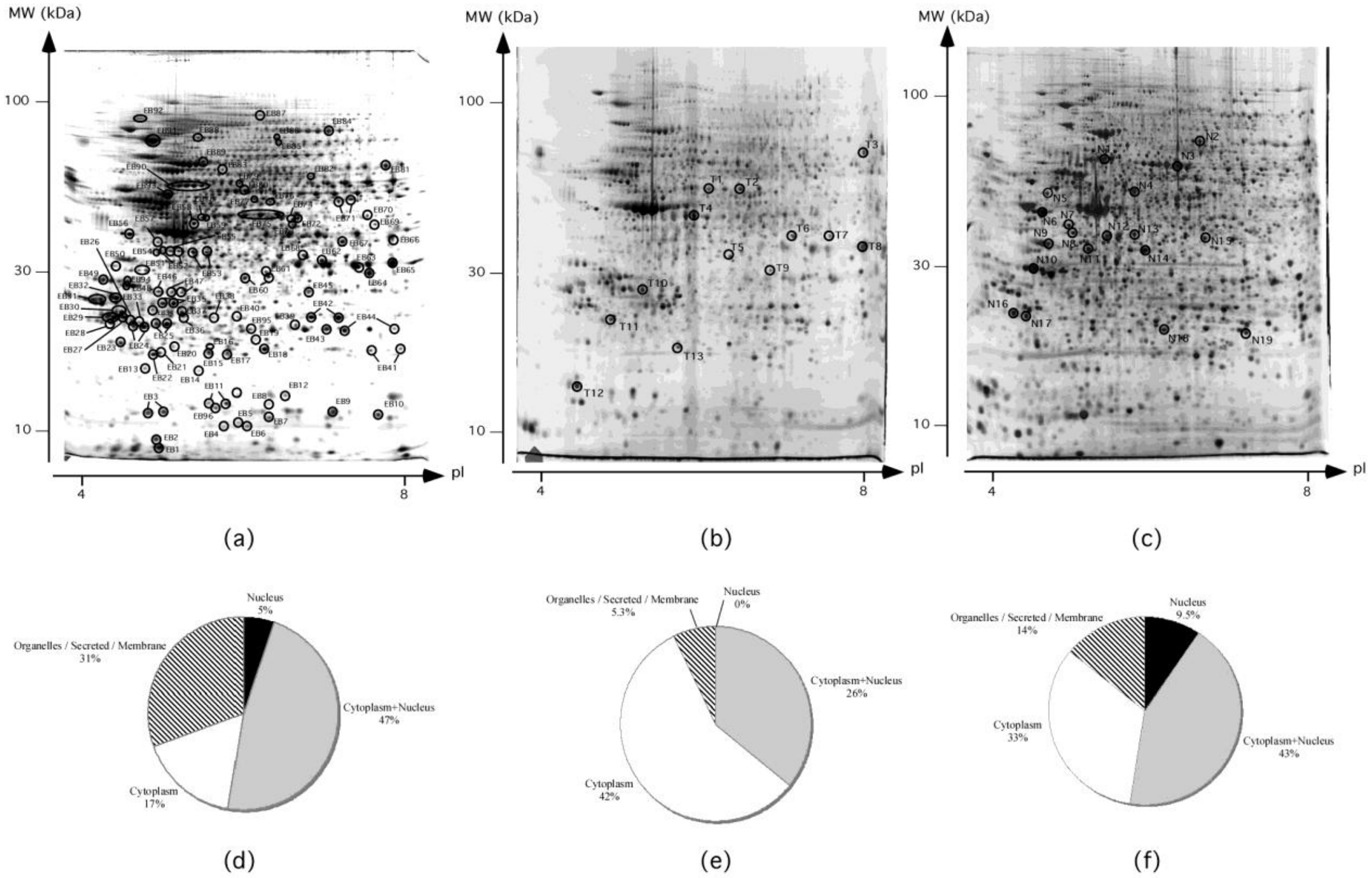

Figure 2

**Figure 2**:

Comparison of the 2D Gel patterns from (a) a J774 total protein extract, (b) a J774 total nuclear protein extract, (c) a J774 NaCl nuclear proteins extract. The comparison has been performed with at least three independent extracts for each method. Proteins systematically present in the total protein extract compared to the nuclear (total or NaCl) extracts have been identified by mass spectrometry (see Table 1). These spots being very abundant because of the very different patterns, we have randomly chosen spots covering most of the area of the gel. The spots systematically enriched in a nuclear extract compared to the other extract have been identified by Mass spectrometry (see Table 1). The piechart shows the ratio of each localization in each extract condition (d) in the total protein extract, (e) in the total nuclear protein extract and (f) in the NaCl nuclear protein extract: white, proteins localized in the cytoplasm, grey, in the cytoplasm and the nucleus, dashed, in the organelles or secreted or in the membrane, and black, in the nucleus.

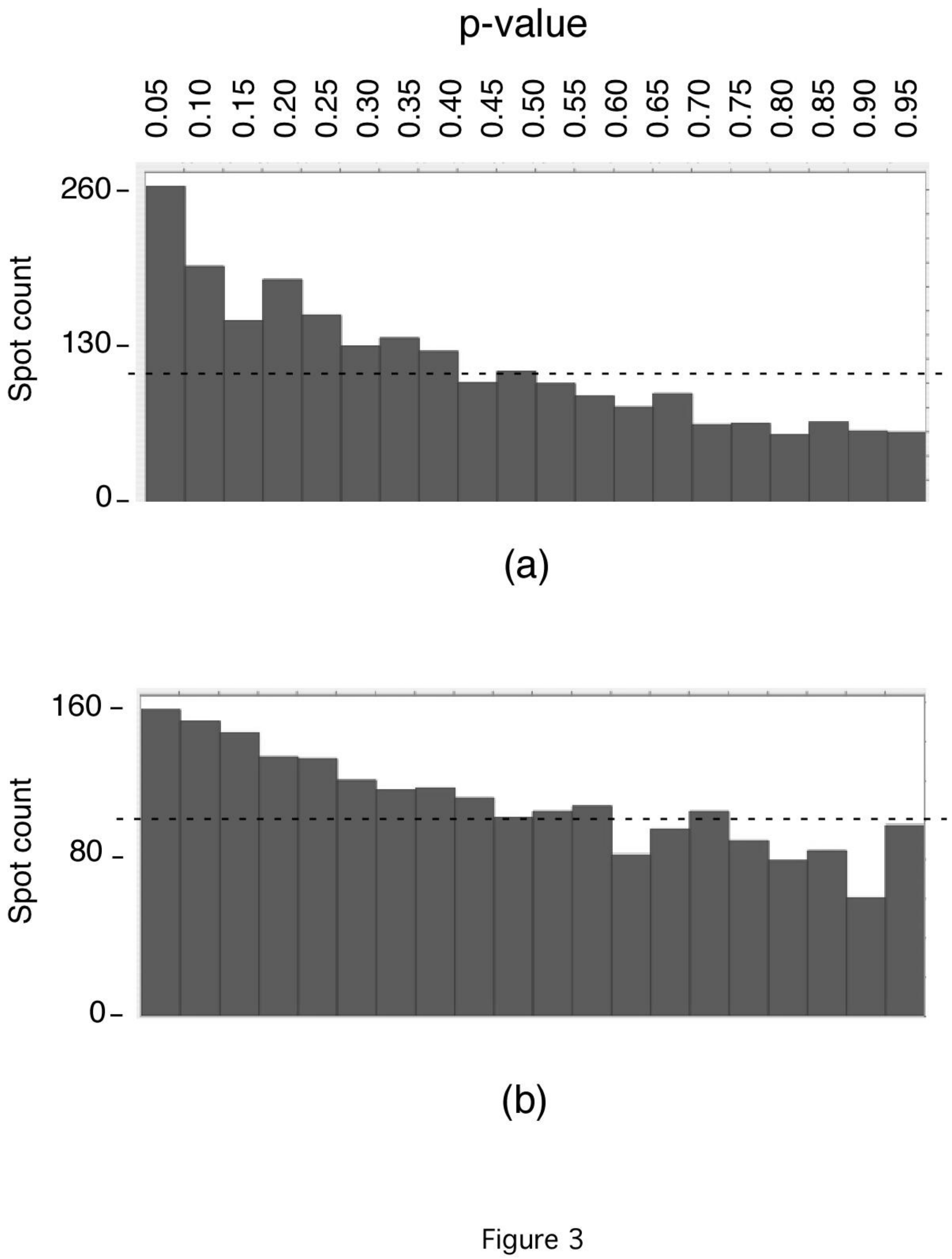

**Figure 3:**

Distribution of the t-tests for all spots detected in the image analysis of the 2D gels (3 independent biological replicates per condition). This allows estimating the proportion of false positives, i.e. spots detected only through random processes, in the selected spots, i.e. those with a t-test lower than 0.05. (a) J774 vs XS52 comparison and (b) J774 vs J774 comparison (null experiment).

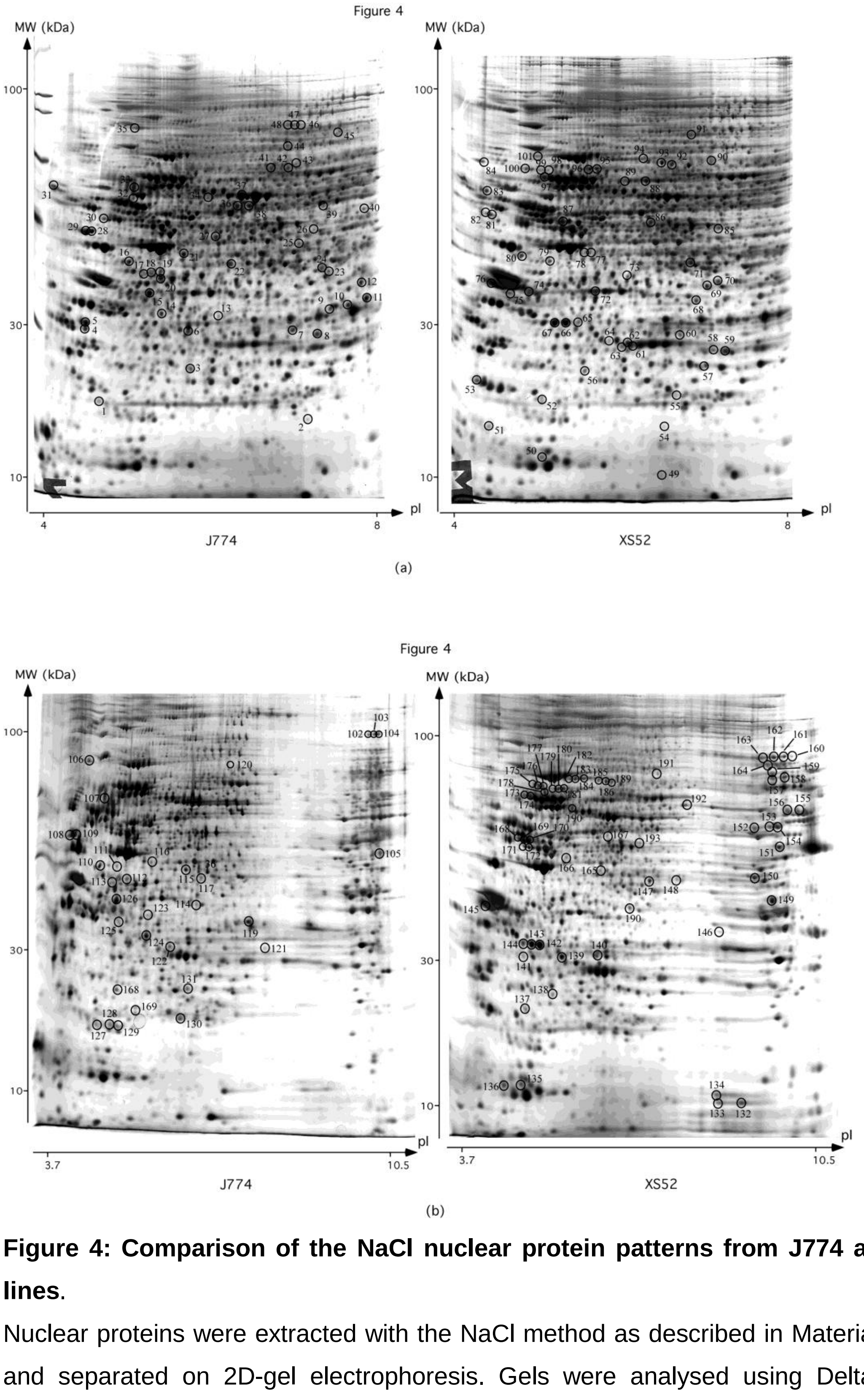

**Figure 4: Comparison of the NaCl nuclear protein patterns from J774 and XS52 cell lines**.

Nuclear proteins were extracted with the NaCl method as described in Materials & Methods and separated on 2D-gel electrophoresis. Gels were analysed using Delta2D software. Circled spots are those differentially expressed by a factor equal or greater than two and a p-value lower than 0.05 in a two-tailed t-test. They have been identified by mass spectrometry see Table 2. (a) pH gradient 4-8 and (b) pH gradient 3.7-10.5.

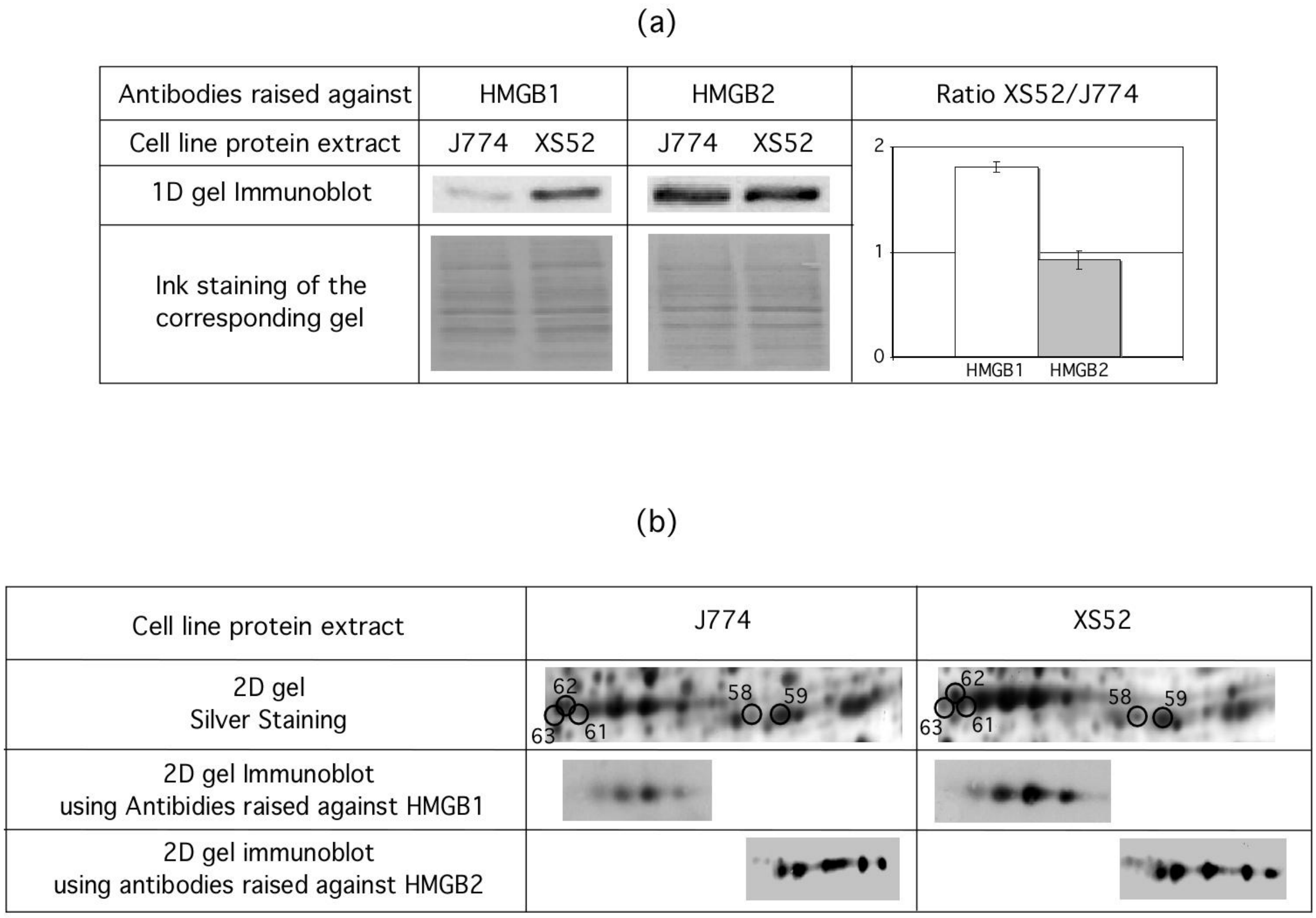

Figure 5

**Figure 5: HMGB1&2 expression patterns**

(a) 1D-gel immunoblotting analysis: Total protein extracts from J774 and XS52 cell lines were separated on 10% SDS-PAGE gel, transferred and then probed with appropriate antibodies raised against HMGB1 or HMGB2. The histogram shows the average ratio of XS52 band relatively to the J774 band, each being normalized to the total amount of proteins quantified with the ink staining. The average is the result of at least three independent extracts (white and grey represent the analysis using antibodies raised against HMGB1 and HMGB2, respectively). (b) 2D-gel immunoblotting analysis: Total protein extracts from J774 and XS52 cell lines were separated on 2D-gel, transferred and then probed with appropriate antibodies raised against HMGB1 or HMGB2. The same part of 2D-gel electrophoresis has been in one hand, silver stained (first line) and, in the other hand, revealed with HMGB1 (second lane) or HMGB2 (third lane) antibodies. The 2D-gel analysis has been repeated at least three times with independent extracts. The spot numbers are the same as Figure 4 and Table 2.

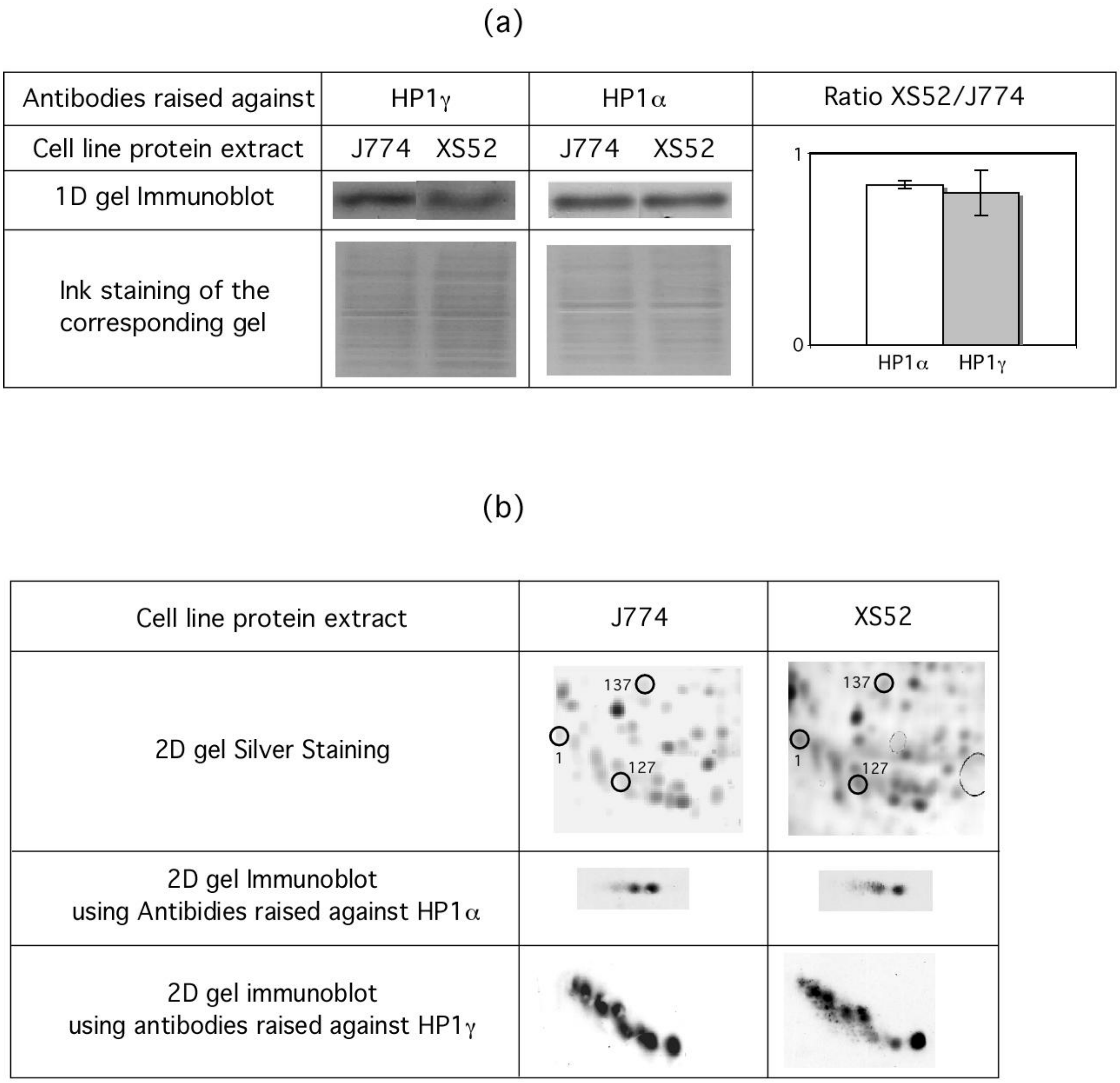

Figure 6

**Figure 6: HP1α, HP1γ expression pattern**

(a) 1D-gel immunoblotting analysis: Total protein extracts from J774 and XS52 cell lines were separated on 10% SDS-PAGE gel, transferred and then probed with appropriate antibodies raised against HP1α or HP1γ. The histogram shows the average ratio of XS52 band relatively to the J774 band, each being normalized to the total amount of proteins quantified with the ink staining. The average is the result of at least three independent extracts (white and grey represent the analysis using antibodies raised against HP1α and HP1γ, respectively). (b) 2D-gel immunoblotting analysis: Total protein extracts from J774 and XS52 cell lines were separated on 2D-gel, transferred and then probed with appropriate antibodies raised against HP1$\alpha$ or HP1$\gamma$. The same part of 2D-gel electrophoresis has been in one hand, silver stained (first line) and, in the other hand, revealed with HP1$\alpha$ (second lane) or HP1$\gamma$ (third lane) antibodies. The 2D-gel analysis has been repeated at least three times with independent extracts. The spot numbers are the same as Figure 4 and Table 2.

**Bibliography**

[1] Desrivieres S, Kuhn K, Muller J, Glaser M, Laria NCP, Kordet J, et al. Comparison of the nuclear proteomes of mammary epithelial cells at different stages of functional differentiation. Proteomics. 2007;7:2019-37.
[2] Gauci S, Veenhoff LM, Heck AJR, Krijgsveld J. Orthogonal Separation Techniques for the Characterization of the Yeast Nuclear Proteome. Journal of Proteome Research. 2009;8:3451-63.
[3] Shen J, Zhu H, Xiang X, Yu Y. Differential Nuclear Proteomes in Response to N-Methyl-N '-nitro-N-nitrosoguanidine Exposure. Journal of Proteome Research. 2009;8:2863-72.
[4] Han B, Stockwin LH, Hancock C, Yu SX, Hollingshead MG, Newton DL. Proteomic Analysis of Nuclei Isolated from Cancer Cell Lines Treated with Indenoisoquinoline NSC 724998, a Novel Topoisomerase I Inhibitor. Journal of Proteome Research.9:4016-27.
[5] Henrich S, Cordwell SJ, Crossett B, Baker MS, Christopherson RI. The nuclear proteome and DNA-binding fraction of human Raji lymphoma cells. Biochimica Et Biophysica Acta-Proteins and Proteomics. 2007;1774:413-32.
[6] Rabilloud T, Pennetier JL, Hibner U, Vincens P, Tarroux P, Rougeon F. Stage Transitions in Lymphocyte-B Differentiation Correlated with Limited Variations in Nuclear Proteins. Proceedings of the National Academy of Sciences of the United States of America. 1991;88:1830-4.
[7] LeStourgeon WM, Beyer AL. The rapid isolation, high-resolution electrophoretic characterization, and purification of nuclear proteins. Methods Cell Biol. 1977;16:387-406.
[8] Chevallet M, Diemer H, Van Dorssealer A, Villiers C, Rabilloud T. Toward a better analysis of secreted proteins: the example of the myeloid cells secretome. Proteomics. 2007;7:1757-70.
[9] Matuo Y, Matsui SI, Nishi N, Wada F, Sandberg AA. Quantitative Solubilization of Nonhistone Chromosomal-Proteins without Denaturation Using Zwitterionic Detergents. Analytical Biochemistry. 1985;150:337-44.
[10] Willard KE, Giometti CS, Anderson NL, Oconnor TE, Anderson NG. Analytical Techniques for Cell-Fractions .26. Two-Dimensional Electrophoretic Analysis of Basic-Proteins Using Phosphatidyl Choline Urea Solubilization. Analytical Biochemistry. 1979;100:289-98.
[11] Neuhoff V, Arold N, Taube D, Ehrhardt W. Improved staining of proteins in polyacrylamide gels including isoelectric focusing gels with clear background at nanogram sensitivity using Coomassie Brilliant Blue G-250 and R-250. Electrophoresis. 1988;9:255-62.
[12] Gianazza E, Celentano F, Magenes S, Ettori C, Righetti PG. Formulations for Immobilized Ph Gradients Including Ph Extremes. Electrophoresis. 1989;10:806-8.
[13] Rabilloud T, Valette C, Lawrence JJ. Sample Application by in-Gel Rehydration Improves the Resolution of 2-Dimensional Electrophoresis with Immobilized Ph Gradients in the First-Dimension. Electrophoresis. 1994;15:1552-8.
[14] Rabilloud T, Adessi C, Giraudel A, Lunardi J. Improvement of the solubilization of proteins in two-dimensional electrophoresis with immobilized pH gradients. Electrophoresis. 1997;18:307-16.
[15] Luche S, Diemer H, Tastet C, Chevallet M, Van Dorsselaer A, Leize-Wagner E, et al. About thiol derivatization and resolution of basic proteins in two-dimensional electrophoresis. Proteomics. 2004;4:551-61.
[16] Tastet C, Lescuyer P, Diemer H, Luche S, van Dorsselaer A, Rabilloud T. A versatile electrophoresis system for the analysis of high- and low-molecular-weight proteins. Electrophoresis. 2003;24:1787-94.

[17] Chevallet M, Luche S, Rabilloud T. Silver staining of proteins in polyacrylamide gels. Nature Protocols. 2006;1:1852-8.
[18] Storey JD, Tibshirani R. Statistical significance for genomewide studies. Proceedings of the National Academy of Sciences of the United States of America. 2003;100:9440-5.
[19] Karp NA, McCormick PS, Russell MR, Lilley KS. Experimental and statistical considerations to avoid false conclusions in proteomics studies using differential in-gel electrophoresis. Molecular & Cellular Proteomics. 2007;6:1354-64.
[20] Gharahdaghi F, Weinberg CR, Meagher DA, Imai BS, Mische SM. Mass spectrometric identification of proteins from silver-stained polyacrylamide gel: A method for the removal of silver ions to enhance sensitivity. Electrophoresis. 1999;20:601-5.
[21] Richert S, Luche S, Chevallet M, Van Dorsselaer A, Leize-Wagner E, Rabilloud T. About the mechanism of interference of silver staining with peptide mass spectrometry. Proteomics. 2004;4:909-16.
[22] Luche S, Lelong C, Diemer H, Van Dorsselaer A, Rabilloud T. Ultrafast coelectrophoretic fluorescent staining of proteins with carbocyanines. Proteomics. 2007;7:3234-44.
[23] Haycock JW. Polyvinylpyrrolidone as a Blocking-Agent in Immunochemical Studies. Analytical Biochemistry. 1993;208:397-9.
[24] Hancock K, Tsang VCW. India Ink Staining of Proteins on Nitrocellulose Paper. Analytical Biochemistry. 1983;133:157-62.
[25] Calogero S, Grassi F, Aguzzi A, Voigtlander T, Ferrier P, Ferrari S, et al. The lack of chromosomal protein Hmg1 does not disrupt cell growth but causes lethal hypoglycaemia in newborn mice. Nature Genetics. 1999;22:276-80.
[26] Nemeth M, Anderson S, Kirby M, Bodine D. Hmgb3 regulates the balance between HSC differentiation and self renewal. Blood. 2005;106:1718.
[27] Jayaraman L, Moorthy NC, Murthy KGK, Manley JL, Bustin M, Prives C. High mobility group protein-1 (HMG-1) is a unique activator of p53. Genes & Development. 1998;12:462-72.
[28] Golob M, Buettner R, Bosserhoff AK. Characterization of a transcription factor binding site, specifically activating MIA transcription in melanoma. Journal of Investigative Dermatology. 2000;115:42-7.
[29] Poser I, Golob M, Buettner R, Bosserhoff AK. Upregulation of HMG1 leads to melanoma inhibitory activity expression in malignant melanoma cells and contributes to their malignancy phenotype. Molecular and Cellular Biology. 2003;23:2991-8.
[30] Sasahira T, Kirita T, Oue N, Bhawal UK, Yamamoto K, Fujii K, et al. High mobility group box-1-inducible melanoma inhibitory activity is associated with nodal metastasis and lymphangiogenesis in oral squamous cell carcinoma. Cancer Science. 2008;99:1806-12.
[31] El Gazzar M, Yoza BK, Chen XP, Garcia BA, Young NL, McCall CE. Chromatin-Specific Remodeling by HMGB1 and Linker Histone H1 Silences Proinflammatory Genes during Endotoxin Tolerance. Molecular and Cellular Biology. 2009;29:1959-71.
[32] Bonaldi T, Talamo F, Scaffidi P, Ferrera D, Porto A, Bachi A, et al. Monocytic cells hyperacetylate chromatin protein HMGB1 to redirect it towards secretion. Embo Journal. 2003;22:5551-60.
[33] Rabbani A, Goodwin GH, Johns EW. Studies on Tissue Specificity of High-Mobility-Group Non-Histone Chromosomal-Proteins from Calf. Biochemical Journal. 1978;173:497-505.
[34] Lomberk G, Wallrath LL, Urrutia R. The Heterochromatin Protein 1 family. Genome Biology. 2006;7.
[35] Grewal SIS, Jia ST. Heterochromatin revisited. Nature Reviews Genetics. 2007;8:35-46.

[36] Lomberk G, Bensi D, Fernandez-Zapico ME, Urrutia R. Evidence for the existence of an HP1-mediated subcode within the histone code. Nature Cell Biology. 2006;8:407-U62.
[37] Leroy G, Weston JT, Zee BM, Young NL, Plazas-Mayorca MD, Garcia BA. Heterochromatin Protein 1 Is Extensively Decorated with Histone Code-like Post-translational Modifications. Molecular & Cellular Proteomics. 2009;8:2432-42.
[38] Takanashi M, Oikawa K, Fujita K, Kudo M, Kinoshita M, Kuroda M. Heterochromatin Protein 1 gamma Epigenetically Regulates Cell Differentiation and Exhibits Potential as a Therapeutic Target for Various Types of Cancers. American Journal of Pathology. 2009;174:309-16.
[39] Petrak J, Ivanek R, Toman O, Cmejla R, Cmejlova J, Vyoral D, et al. Deja vu in proteomics. A hit parade of repeatedly identified differentially expressed proteins. Proteomics. 2008;8:1744-9.
[40] Wang P, Bouwman FG, Mariman EC. Generally detected proteins in comparative proteomics--a matter of cellular stress response? Proteomics. 2009;9:2955-66.
[41] Gronow M, Griffith.G. Rapid Isolation and Separation of Non-Histone Proteins of Rat Liver Nuclei. Febs Letters. 1971;15:340-&.
[42] Caruccio L, Banerjee R. An efficient method for simultaneous isolation of biologically active transcription factors and DNA. Journal of Immunological Methods. 1999;230:1-10.
[43] Horton P, Park KJ, Obayashi T, Nakai K. Protein subcellular localization prediction with WOLF PSORT.  Proceedings of the 4th Asia-Pacific Bioinformatics Conference2006. p. 39-48.